# Evolution of the Earth's Atmosphere during Late Veneer Accretion

Catriona A. Sinclair,[1]* Mark C. Wyatt,[1] Alessandro Morbidelli,[2] David Nesvorný[3]

[1]*Institute of Astronomy, Madingley Road, Cambridge CB3 0HA, UK*
[2]*Laboratoire Lagrange, Université Côte d'Azur, Ovservatoire de la Université Côte d'Azur, CNRS*
[3]*Department of Space Studies, Southwest Research Institute, 1050 Walnut St., Suite 300, Boulder, CO, 80302, USA*

**ABSTRACT**
Recent advances in our understanding of the dynamical history of the Solar system have altered the inferred bombardment history of the Earth during accretion of the Late Veneer, after the Moon-forming impact. We investigate how the bombardment by planetesimals left-over from the terrestrial planet region after terrestrial planet formation, as well as asteroids and comets, affects the evolution of Earth's early atmosphere. We develop a new statistical code of stochastic bombardment for atmosphere evolution, combining prescriptions for atmosphere loss and volatile delivery derived from hydrodynamic simulations and theory with results from dynamical modelling of realistic populations of impactors. We find that for an initially Earth-like atmosphere impacts cause moderate atmospheric erosion with stochastic delivery of large asteroids giving substantial growth ($\times 10$) in a few % of cases. The exact change in atmosphere mass is inherently stochastic and dependent on the dynamics of the left-over planetesimals. We also consider the dependence on unknowns including the impactor volatile content, finding that the atmosphere is typically completely stripped by especially dry left-over planetesimals ($< 0.02$ % volatiles). Remarkably, for a wide range of initial atmosphere masses and compositions, the atmosphere converges towards similar final masses and compositions, i.e. initially low mass atmospheres grow whereas massive atmospheres deplete. While the final properties are sensitive to the assumed impactor properties, the resulting atmosphere mass is close to that of current Earth. The exception to this is that a large initial atmosphere cannot be eroded to the current mass unless the atmosphere was initially primordial in composition.

## 1 INTRODUCTION

We are observing ever-increasing numbers of exoplanets (Kaltenegger 2017; Defrère et al. 2018), with precision such that it is now possible to carry out atmospheric characterisation of terrestrial planets orbiting within the habitable zone of their host star (de Wit et al. 2018). In light of this, it is necessary to understand the origin and evolution of these atmospheres. This in turn requires understanding of the processes that govern both atmospheric loss and growth. The present day geochemical inventory of the atmospheres of Earth and other Solar system bodies can give insight into their sources, as well as the potential loss and delivery mechanisms that have been acting on them over their history. Earth's atmosphere has undergone substantial evolution over its lifetime as a result of many processes acting on different spatial and temporal scales. Due to the comparatively abundant observational constraints provided by the Earth, it is an important test case for understanding these processes.

Growth of Earth's atmosphere might have occurred through accretion of primordial gases from the pre-Solar nebula, mantle outgassing from magma oceans or volcanic outgassing while the atmosphere might have been eroded as a result of impacts during the final stages of planet formation, hydrodynamic escape driven by extreme-UV flux from the active young Sun, or mantle ingassing by a magma ocean. Each of these processes should leave observable traces in the geochemistry of the atmosphere and solid Earth. For example, the relative abundances of volatiles in Earth's atmosphere are not chondritic (Halliday 2013), yet the isotopic ratios are (Marty 2012), implying that the atmosphere has undergone bulk removal rather than hydrodynamic escape which is predicted to preferentially remove lighter isotopes and so should result in isotope fractionation (Schlichting & Mukhopadhyay 2018). There is isotopic evidence from Earth's mantle for a series of giant impact induced magma ocean phases and associated episodes of atmosphere loss. However, outgassing during these events (even in combination with hydrodynamic loss) cannot explain the mantle and atmosphere noble gas observations (Schlichting & Mukhopadhyay 2018). These observations lead to the conclusion that atmospheric erosion caused by impacts is likely to have played an important role in the early evolution of Earth's atmosphere. Furthermore, the delivery of volatiles through impacts is also likely to have been significant in the evolution of Earth's atmosphere, and is a topic that is far from resolved (see for example Chyba 1990; Elkins-Tanton et al. 2011; Halliday 2013; Marty et al. 2016; Zahnle et al. 2019a).

---

* E-mail: cas213@cam.ac.uk

In order to understand the evolution of the Earth's atmosphere as a result of impacts, it is necessary to understand both the properties of the impactors (size, velocity, composition, impact flux, etc.), which may vary over time, and the effect of each impact on the atmosphere. Within the context of the Solar system these impactors can be broadly categorised into three populations, comets, asteroids and planetesimals left-over from the terrestrial planet region after terrestrial planet formation (hereafter referred to as left-over planetesimals), the latter of which has not yet been studied in the context of their effect on the Earth's atmosphere. Furthermore, the initial atmosphere on the Earth is also an important factor in determining the atmospheric evolution resulting from impacts, with the same impactor population theoretically capable of causing growth of an initially small atmosphere and loss of an initially large atmosphere (Wyatt et al. 2019). This can lead to an equilibrium solution, whereby the atmosphere mass is maintained at a constant value by impacts, as discussed in Schlichting et al. (2015). This is a significant uncertainty, with no direct observational constraints for the mass and composition of the Earth's early atmosphere, and only scant proxy observations through preserved zircons (Harrison et al. 2017).

The outcome of an impact on the atmosphere will depend on the energy of the impactor: whether the impactor reaches the surface intact, and if it does whether it causes global effects. The effect of impacts on planetary atmospheres has long been studied, with the first models (Melosh & Vickery 1989; Vickery & Melosh 1990; Ahrens 1993) using the vapour plume expansion model of Zel'dovich & Raizer (1967). This was extended to consider the evolution of atmospheres during an extended period of bombardment in Zahnle et al. (1992), while giant impacts were studied in Genda & Abe (2003). Detailed simulations of atmosphere loss and delivery were carried out by Svetsov (2007) and Shuvalov (2009) considering a range of impactor parameters including density, velocity and impact angle. A theoretical framework within which to understand these simulation results was given in Schlichting et al. (2015), in which the effect of non-local atmosphere mass loss caused by large impacts was also investigated. More recent studies have also been performed using smoothed-particle hydrodynamics to model the outcomes of giant impacts (Kegerreis et al. 2020).

The evolution of Solar system atmospheres as a result of impacts was first considered in a comparative study between Titan, Ganymede and Callisto by Zahnle et al. (1992), who found that accretion of cometary bodies should result in atmospheric erosion on Ganymede and Callisto but growth on Titan. In Pham et al. (2011) the authors attempted to explain the differences between the present day atmospheres of Earth, Venus and Mars through asteroid and comet impacts, concluding that impacts could erode Mars' atmosphere but deliver volatiles to Venus and Earth. In de Niem et al. (2012) the inherent stochasticity of impacts during a period of heavy bombardment was studied by implementing prescriptions for atmospheric erosion and impactor accretion based on the results of Svetsov (2000). They considered the effect of varying the impactor compositions (through the ratio of asteroid- to comet-like impactors) and the initial atmospheric pressure, finding that impacts by such populations should result in atmospheric growth for both Earth and Mars. The effect of impacts on exoplanet atmospheres was studied for the first time in Kral et al. (2018), which investigated volatile delivery and atmosphere erosion of the primordial atmospheres of the TRAPPIST-1 planets by comets. In Wyatt et al. (2019), an analytical model for the evolution of an atmosphere due to impacts that can both remove atmosphere mass and deliver volatiles, based on the prescription of Shuvalov (2009), was developed. This model quantifies how the growth or loss of an atmosphere depends sensitively on the assumptions made about the impact velocity, and the composition and size distribution of the impactors. When applied to the Earth this showed that either growth or loss were possible, but that a detailed consideration of the impactor population from which reliable conclusions could be drawn was left to a later study (i.e. this paper).

This picture of the Earth undergoing an extended period of bombardment following formation is supported by more detailed considerations of the dynamical evolution of the Solar system. In this paper we will consider this to have occurred within the framework of the Nice model (Tsiganis et al. 2005; Gomes et al. 2005; Morbidelli et al. 2005), which describes the migration of the giant planets from nearly circular, compact, co-planar orbits surrounded by a primordial disk of small icy bodies, onto their present day orbits. The initial conditions for the Nice model can be explained by the "Grand Tack" model (Walsh et al. 2011), which describes the inward then outward resonant migration of Jupiter and Saturn within the protoplanetary disk. The Nice model has been further extended, to include encounters between the giant planets in the "Jumping Jupiter" model, (Morbidelli et al. 2010) in which Jupiter undergoes a number of close encounters with an ice giant. These models have continued to undergo development over the past decade, with the giant planet instability modelled in detail in Nesvorný & Morbidelli (2012), and terrestrial planet bombardment studied in Nesvorný et al. (2017a); Morbidelli et al. (2018). These models lead to the conclusion that rapid migration of the giant planets during the instability results in a period of intense bombardment onto the terrestrial planets, delivering the Late Veneer to Earth in the so-called accretion tail (Morbidelli et al. 2018).

Direct evidence of these early impacts on Earth in the form of craters does not exist due to the processing of the crust by tectonic activity, with the oldest known impact crater only 2.2 Gyr old (Erickson et al. 2020). The lunar cratering record provides constraints on the impact history experienced by the Moon and Earth, as do measurements of mantle abundances of highly siderophile elements (HSEs). Due to their high affinity for metals HSEs are preferentially sequestered into the core during parent body differentiation, consequently, excess HSEs measured in the crust of a body is believed to record the amount of material accreted since core formation. HSE abundances in the mantles of Earth, Moon and Mars are all in chondritic proportions, which supports the argument that their higher than expected HSE abundances are due to delivery in a "Late Veneer" rather than inefficient core formation. Mantle HSE measurements imply that approximately $(0.5 \pm 0.2)$ wt. % of Earth's mass ($3 \times 10^{22}$ kg) was delivered in the Late Veneer, and approximately 0.025 wt % of the Moon's mass ($2 \times 10^{19}$ kg) was delivered in the same manner (Day et al. 2007; Day & Walker 2015; Day et al. 2016). These mass estimates are in agreement with those obtained through Tungsten isotope measurements (Willbold et al. 2015; Touboul et al. 2015).

The accreted masses for the Earth and the Moon imply that the Earth accreted $\sim 10^3$ times more mass that the Moon, despite having a gravitational cross section only 20 times larger. There have been several explanations proposed to resolve this discrepancy, such as a size distribution favouring the largest impactors (Bottke et al. 2010; Genda et al. 2017) (invoking stochastic accretion of the largest impactors), or a size distribution dominated by small bodies (Schlichting et al. 2012) (invoking the difference in gravitational focusing between the Earth and the Moon). Alternatively a combination of lower impactor retention for the Moon (Zhu et al. 2019) and sulfide segregation of HSEs into the core during lunar magma ocean crystallisation and overturn could explain this discrepancy (Elkins-Tanton et al. 2011; Rubie et al. 2016). This final scenario, discussed in detail in Morbidelli et al. (2018), has been a major recent advance in the field, successfully explaining both the HSE constraints on the Earth and the Moon and the lunar cratering record within the context of the most recent numerical simulations of the fluxes of comets, asteroids and left-over planetesimals.

Given these significant recent advances in our understanding of the accretion tail hypothesis and thus the impact chronology of the Earth, in combination with our improved understanding of the atmospheric effects caused by individual impacts it is timely to revisit conclusions made regarding the evolution of Earth's atmosphere in previous work on this subject. Indeed, previous investigations of the evolution of Earth's atmosphere have not considered this new model for the impact history of the Earth; Pham et al. (2011) made no assumptions regarding the dynamical history of the Solar system, while de Niem et al. (2012) sampled their impactor velocities from a distribution based on the Nice model as it was at that time (Gomes et al. 2005; Morbidelli et al. 2010). Thus, this paper combines the most recent results of dynamical modelling (Nesvorný et al. 2017a,b; Morbidelli et al. 2018) assuming that the impact rate onto the terrestrial planets has been declining since planet formation concluded, with prescriptions for the effect of impacts on Earth's atmosphere, to investigate the outcome of impacts by asteroids, comets and left-over planetesimals, building on the analytical model presented in Wyatt et al. (2019). We also investigate how challenging the assumptions made about the inputs to our model may impact the results we obtain: We account for uncertainties in the dynamical history of the Solar system, using multiple dynamical simulations to calculate the impact fluxes and velocities, as well as the uncertainties in the relative contributions of different impactor populations and the compositions of these populations. We also consider variation in the initial conditions for the Earth's atmosphere, which is not well constrained by observations.

This paper is structured as follows. First the numerical code is presented in its most general form in §2. The prescriptions for the outcome of a single impact that are incorporated into our numerical code are presented in §3, for the effects of both cratering impacts and large impacts that can have non-local effects. The code is tested in §4 against the analytical model of Wyatt et al. (2019), and the inclusion of effects resulting from stochasticity, the time evolution of the planet properties and large impacts are discussed. The inputs used for our application of the code to Earth specifically are motivated in §5. The results of the code are then presented in §6, and the dependence of the conclusions drawn on assumptions made about the impactor compositions, and dynamics is discussed. A discussion of the implications of these results for water delivery, as well as potential other sources of atmosphere mass loss and volatile delivery, and a comparison to previous works is presented in §7. The conclusions of this work are then summarised in §8.

## 2 A NUMERICAL ATMOSPHERIC EVOLUTION CODE

### 2.1 General overview

We have developed a statistical code of stochastic bombardment to model the evolution of a terrestrial planet's atmosphere as it undergoes impacts. The code is designed to be modular, with the choice of impactor populations, prescription for the outcome of a given impact, and initial system parameters able to be independently specified. In the following explanation it is assumed that given the atmosphere, planet and impactor properties, it is possible to calculate two properties:

(i) the atmosphere mass removed by each impactor relative to the impactor mass $\frac{m_{\text{atmloss}}}{m_{\text{imp}}}$,

(ii) the impactor mass fraction retained after an impact $\frac{m_{\text{impacc}}}{m_{\text{imp}}}$.

The operation of the code does not depend on the method used to calculate these values. The method we use is discussed in §3. Given a specified population of impactors, the code will evolve the atmosphere through time, tracking the total atmosphere mass, planet mass and atmosphere composition (through the fraction of the total mass that has been delivered by each impactor population). The mean molecular weight, or the number fraction in the atmosphere of the volatile elements H, C, N, O, and S can be calculated from the impactor properties and atmosphere mass fractions. These abundances could in principle be used as input to a chemical evolution model.

#### 2.1.1 Inputs

The code takes as inputs a series of parameters describing the initial conditions of the planet and its atmosphere, and the population of impactors. For the planet, the semi-major axis ($a_{\text{pl}}$), mass ($M_{\text{pl}}$), and bulk density ($\rho_{\text{pl}}$) must be given, as must the mass ($M_*$) and luminosity ($L_*$) of the host star. The initial atmosphere is described in terms of its mass ($m_0$), bulk mean molecular weight ($\mu_0$) and bulk abundances of the specified volatile elements.

It is possible to specify any number of impactor populations, each characterised by an associated bulk density, volatile fraction, composition, size distribution, impact flux and distribution of impact velocities. The bulk density ($\rho_k$), volatile fraction ($x_{v,k}$)[1] and mean molecular weight ($\mu_k$) of the $k$-th impactor population are given as single numbers. The composition of the impactor can also be specified, as the fraction by number of the volatile elements introduced above ($H_k$, $C_k$, $N_k$, $O_k$, and $S_k$). This composition, if specified, will determine the value of $\mu$ for the impactor volatiles. Both the size and impact velocity distributions can be time dependent. The size distribution is specified as the number fraction of impactors, in $N_{\text{size}}$ log spaced size bins between a minimum ($D_{\min}$) and maximum size ($D_{\max}$). The number fraction of objects in the $i$-th size bin, with size $D_i$ (and mass $m_{\text{imp}}(D_i, \rho_k)$), at time $t$, is written here as $f_{\text{N},i,k}(t)$. The impact velocity distribution is specified in a similar manner. The velocity bins are log spaced between a minimum ($v_{\min}$) and maximum ($v_{\max}$) velocity, with $N_{\text{vel}}$ bins in total. The number fraction of objects in the $j$-th velocity bin, with velocity $v_j$ at time $t$ is written here as $f_{\text{v},j,k}(t)$. The final input is $R_k(t)$, the total impactor flux rate of the $k$-th impactor population of all impactor masses as a function of time.

---

[1] This is referred to by the symbol $p_v$ in Wyatt et al. (2019), but to avoid any possible confusion with albedo we use $x_v$ here instead

*2.1.2 Code method*

Due to the range in impactor parameters that we consider (size, velocity, composition) it is not computationally feasible to update the atmosphere properties separately for each individual impactor. Instead, we consider discrete time steps, within which we start from knowledge of the atmosphere and planet properties in the previous time-step (at time $t$), as well as the impactor properties introduced above. In order to reduce computation time, the code makes use of an adaptive time step, discussed further in §4.1.

For each time step ($\Delta t$), we consider the effect of each impactor population in the following manner. The number of impactors $N_{i,j,k}(t)$ with size $D_i$, velocity $v_j$ in the $k$-th population is drawn from a Poisson distribution, with the parameter $\lambda$ (average number of impacts per time interval)

$$\lambda_{i,j,k} = R_k(t)\, f_{\text{N }i,k}(t)\, f_{\text{v }j,k}(t)\, \Delta t. \quad (1)$$

For this work, the values of $\left(\frac{m_{\text{atmloss}}}{m_{\text{imp}}}\right)_{i,j,k}$ and $\left(\frac{m_{\text{impacc}}}{m_{\text{imp}}}\right)_{i,j,k}$ are calculated using the following approach:

(i) The effect of cratering impacts is accounted for using the chosen prescription (see §3.1).
(ii) The atmosphere mass loss caused by a single impactor is bounded from above by the polar cap mass (equation 16).
(iii) The effects of non-cratering impacts are included (see §3.2).

The total atmosphere mass loss caused by all impactors in that population is then calculated by summing over the size and velocity bins,

$$m_{\text{atmloss},k} = \sum_{i=1}^{N_{\text{size}}} \sum_{j=1}^{N_{\text{vel}}} \left[ N_{i,j,k}(t) m_{\text{imp}}(D_i, \rho_k) \left(\frac{m_{\text{atmloss}}}{m_{\text{imp}}}\right)_{i,j,k} \right]. \quad (2)$$

The total mass of impactor material accreted is then calculated in a similar fashion,

$$m_{\text{impacc},k} = \sum_{i=1}^{N_{\text{size}}} \sum_{j=1}^{N_{\text{vel}}} \left[ N_{i,j,k}(t) m_{\text{imp}}(D_i, \rho_k) \left(\frac{m_{\text{impacc}}}{m_{\text{imp}}}\right)_{i,j,k} \right]. \quad (3)$$

The volatile content $x_{\text{v},k}$ is defined such that it corresponds to the volatiles that will end up in the atmosphere, so it is then used to calculate the mass that is accreted as a solid onto the planet mass, and the mass that is added to the atmosphere. With all these values calculated, the planet mass can then be updated for the next time step as

$$M_{\text{pl}}(t+\Delta t) = M_{\text{pl}}(t) + \sum_{k=1}^{N_{\text{comp}}} \left[m_{\text{impacc},k}\left(1 - x_{\text{v},k}\right)\right]. \quad (4)$$

The atmosphere mass is then calculated in a two step process. On the assumption that atmospheric mass removal occurs before the volatiles delivered are released, the total atmospheric mass loss is calculated from the sum of the contribution by each impactor type, and used to calculate the intermediate atmospheric mass

$$m_{\text{mid}} = m(t) - \sum_{k=1}^{N_{\text{comp}}} \left(m_{\text{atmloss},k}\right). \quad (5)$$

If this value is negative a time step warning flag is raised, since this is not physical, and as discussed in §4.1 typically only occurs if the atmosphere mass is particularly low or the impactor is particularly extreme. If this is the case, then $m_{\text{mid}}$ is set to be zero[2]. The new atmosphere mass is then calculated from the sum of the masses of volatiles delivered by each impactor type as

$$m(t+\Delta t) = m_{\text{mid}} + \sum_{k=1}^{N_{\text{comp}}} \left[m_{\text{impacc},k}\, x_{\text{v},k}\right]. \quad (6)$$

For comparison with the analytic models of Wyatt et al. (2019), we also record the value of $f_{\text{v}}$, the ratio of atmosphere mass gain and mass loss rates

$$f_{\text{v}}(t) = \frac{\dot{m}^+}{\dot{m}^-} = \frac{\sum_{k=1}^{N_{\text{comp}}} \left[m_{\text{impacc},k}\, x_{\text{v},k}\right]}{\min\left(\sum_{k=1}^{N_{\text{comp}}} \left[m_{\text{atmloss},k}\right],\ m(t)\right)}. \quad (7)$$

This comparison is discussed further in §4.4. The change in $\mu$ due to the impacts in that time-step are calculated, through

$$\mu(t+\Delta t) = \frac{\left(m_{\text{mid}}\mu(t) + \sum_{k=1}^{N_{\text{comp}}} \left[m_{\text{impacc},k}\, x_{\text{v},k}\, \mu_{\text{imp},k}\right]\right)}{m(t+\Delta t)}. \quad (8)$$

The code tracks the atmosphere mass at each time that has been delivered by each impactor population ($m_k(t)$), which is calculated through

$$m_k(t+\Delta t) = \left(m_k(t) - f_k(t) \sum_{k=1}^{N_{\text{comp}}} \left[m_{\text{atmloss},k}\right] + m_{\text{impacc},k}\, x_{\text{v},k}\right). \quad (9)$$

This is used to calculate the corresponding atmosphere mass fraction $f_k(t) \equiv \frac{m_k(t)}{m(t)}$.

Assuming that the initial volatile content of the the atmosphere is known, $X_{\text{atm}}(0)$, where $X \equiv [\text{H, C, N, O, S}]$, and the volatile content of each of the impacting populations ($X_{\text{imp}k}$) is also known, then this fraction can be used to calculate the evolution of the relative abundances of the volatile species in the atmosphere through

$$X(t) = \sum_{k=1}^{N_{\text{comp}}} \left[f_k(t) X_{\text{imp}k}\right] + \left(1 - \sum_{k=1}^{N_{\text{comp}}} \left[f_k(t)\right]\right) X_{\text{atm}}(0). \quad (10)$$

---

[2] For numerical reasons it is actually set to an arbitrarily small number, typically $10^{-55}$ M$_\oplus$

## 3 PRESCRIPTION FOR THE EFFECT OF AN IMPACT

The outcome of an impact on the atmosphere will depend on the energy of the impactor, whether the impactor reaches the surface intact, and if it does whether it causes global effects. The numerical code combines multiple prescriptions for the outcome of an impact in an attempt to cover as much of the parameter space spanned by potential impactors as possible. Cratering impacts, in which the impactor reaches the ground intact but is not so large as to cause a global shock wave through the planet, are described by the prescriptions presented in Shuvalov (2009), and have been incorporated into an analytic model described in Wyatt et al. (2019). The analytic prescriptions from Shuvalov (2009) are fits to the angle averaged results from simulations of left-over planetesimals with a range of sizes ($D = 1 - 30$ km), impact velocities ($v_{\mathrm{imp}} = 10 - 70$ km s$^{-1}$) and impactor densities ($\rho_{\mathrm{imp}}$), undergoing cratering impacts on an Earth-like planet.

The atmosphere mass changes resulting from a single impact are presented in terms of a single parameter, the erosional efficiency ($\eta$). This is a function of the impactor properties ($D$, $\rho_{\mathrm{imp}}$ and $v_{\mathrm{imp}}$) and planet properties (escape velocity $v_{\mathrm{esc}}$, density $\rho_{\mathrm{pl}}$, atmosphere scale height $H$, and atmospheric density at the base of the atmosphere $\rho_0$) through

$$\eta = \left(\frac{D}{H}\right)^3 \left[\left(\frac{v_{\mathrm{imp}}}{v_{\mathrm{esc}}}\right)^2 - 1\right] \left[\frac{\rho_{\mathrm{imp}}\rho_{\mathrm{pl}}}{\rho_0(\rho_{\mathrm{imp}} + \rho_{\mathrm{pl}})}\right]. \quad (11)$$

The prescriptions can be extended to parameters beyond the scope of the initial simulations, and this can be physically justified (Schlichting et al. 2015). However, this should be done with caution, as if impactors do not reach the ground with the energy required to cause a cratering event then different physical processes dominate the outcome. Impacts of this kind, where the impactors may be slowed sufficiently to cause aerial bursts or fragment during their flight are not considered here, but merit future investigation. The method by which cratering impacts are incorporated into the code is described in §3.1. Large impactors, that can remove all the atmosphere in the vicinity of their impact site, as well as cause non-local atmosphere loss through global ground motion, are described in Schlichting et al. (2015) and Yalinewich & Schlichting (2019), and their inclusion is discussed in §3.2. The combination of these prescriptions is described in §3.3, and our extrapolation into the regime where the atmosphere mass becomes very small, called the "airless limit" is discussed in §3.4.

### 3.1 Cratering impacts

Assuming an isothermal, ideal atmosphere at temperature $T = 278\, L_*^{0.25}\, a_{\mathrm{pl}}^{-0.5}$ K, where $L_*$ is the luminosity of the host star in units of L$_\odot$ and $a_{\mathrm{pl}}$ is the semi-major axis in au, the scale height is $H = 0.73 \times 10^6\, L_*^{0.25}\, a_{\mathrm{pl}}^{-0.5}\, M_{\mathrm{pl}}^{-1/3}\, \rho_{\mathrm{pl}}^{-2/3}\, \mu^{-1}$ m, where $\rho_{\mathrm{pl}}$ is in g cm$^{-3}$ and $M_{\mathrm{pl}}$ is the planet mass in M$_\oplus$. This can be combined with equation 11, further assuming that $H \ll R_{\mathrm{pl}}$ so that the total atmosphere mass can be approximated by $m \equiv \delta M_{\mathrm{pl}} \approx 4\pi R_{\mathrm{pl}}^2 H \rho_0$, to give (Wyatt et al. 2019)

$$\eta = 0.5 \times 10^{-18} L_*^{-0.5} a_{\mathrm{pl}} M_{\mathrm{pl}}^{1/3} \rho_{\mathrm{pl}}^{5/3} \delta^{-1} \mu^2 D^3 \frac{\left[\left(\frac{v_{\mathrm{imp}}}{v_{\mathrm{esc}}}\right)^2 - 1\right]}{\left(1 + \frac{\rho_{\mathrm{pl}}}{\rho_{\mathrm{imp}}}\right)}. \quad (12)$$

The prescription from Shuvalov (2009) for cratering impacts gives the fractional atmosphere mass lost due to a single impact by a body with mass $m_{\mathrm{imp}} = \frac{\pi}{6}\rho_{\mathrm{imp}}D^3$ to be

$$\frac{m_{\mathrm{atmloss}}}{m_{\mathrm{imp}}} = \left[\left(\frac{v_{\mathrm{imp}}}{v_{\mathrm{esc}}}\right)^2 - 1\right]\chi_{\mathrm{a}}(\eta), \quad (13)$$

where

$$\begin{aligned}\log\left(\chi_{\mathrm{a}}(\eta)\right) = &-6.375 + 5.239\bigl(\log(\eta)\bigr) - 2.121\bigl(\log(\eta)\bigr)^2 \\ &+ 0.397\bigl(\log(\eta)\bigr)^3 - 0.037\bigl(\log(\eta)\bigr)^4 \\ &+ 0.0013\bigl(\log(\eta)\bigr)^5.\end{aligned}$$
(14)

The prescription for atmosphere loss given by equation 14 is not appropriate for large values of $\eta$ (the most energetic impactors). The simulations performed in Shuvalov (2009) cover only up to $\eta \sim 10^6 - 10^7$. Naively applying the prescription for atmosphere mass loss above results in predictions that absolute atmospheric mass loss will start to decrease with increasing impactor mass for values of $\eta > 10^{6.3}$. This is not physically realistic, so we choose to modify the prescription for $\chi_a$ for values of $\eta > 10^6$. This modification takes the form of a power law, fit to $\chi_a(10^4 \leqslant \eta \leqslant 10^6)$, applied to $\eta > 10^6$ with a correction to avoid a discontinuity in $\chi_a$ at $\eta = 10^6$

$$\log\left(\chi_a(\eta > 10^6)\right) = -0.6438\eta + 0.4746. \quad (15)$$

This results in absolute atmosphere mass losses that increase with increasing impactor mass until the polar cap limit is reached, in line with the theoretical framework presented in Schlichting et al. (2015). The polar cap limit arises from the constraint that the maximum atmosphere mass that can be ejected by a single cratering impact cannot be greater than the mass of the atmosphere contained in the polar cap above the impact site

$$m_{\mathrm{max}} = m_{\mathrm{cap}} = 2\pi\rho_0 H^2 R_{\mathrm{pl}}. \quad (16)$$

This is incorporated into the prescription given by equation 13 as an upper bound on the value of $m_{\mathrm{atmloss}}$ calculated for a single impactor of each size, before the contributions of all impactors of that size are combined.

The fractional impactor mass accreted by the planet due to a single impact is given by

$$\frac{m_{\mathrm{impacc}}}{m_{\mathrm{imp}}} = \begin{cases} 1 & \eta \leqslant 10, \\ 1 - \chi_{\mathrm{pr}}(\eta) & 10 \leqslant \eta < 1000, \\ 1 - \chi_{\mathrm{pr}}(\eta = 1000) & 1000 < \eta, \end{cases} \quad (17)$$

where

$$\chi_{\mathrm{pr}}(\eta) = \min\left[0.07\left(\frac{\rho_{\mathrm{pl}}}{\rho_{\mathrm{imp}}}\right)\left(\frac{v_{\mathrm{imp}}}{v_{\mathrm{esc}}}\right)\bigl(\log(\eta) - 1\bigr), 1\right]. \quad (18)$$

### 3.2 Large impacts

For the most energetic (the largest and fastest) impactors, the prescription for atmosphere loss given by equation 13 is not a physically complete description, as violent impacts can cause a shock wave to propagate through the planet, causing non-local atmosphere loss by accelerating regions of the atmosphere on the opposite side of the planet beyond the escape velocity. A prescription for this effect is given by Schlichting et al. (2015) (for the isothermal atmospheres considered here)

$$\frac{m_{\mathrm{atmloss,GI}}}{m_{\mathrm{imp}}} = \delta\frac{v_{\mathrm{imp}}}{v_{\mathrm{esc}}}\left[0.4 + 1.4x - 0.8x^2\right], \quad (19)$$

where $x \equiv \left(\frac{v_{\rm imp}}{v_{\rm esc}}\right)\left(\frac{m_{\rm imp}}{M_{\rm pl}}\right)$.

The combination of the Shuvalov (2009) cratering and the Schlichting et al. (2015) (S15) prescriptions described in §3 is not the only available method for parameterising the atmosphere loss caused by impacts. One alternative is the prescription presented in Kegerreis et al. (2020) (K20), which parameterises atmosphere loss as a function of specific impact energy

$$Q = (1-b)^2 \frac{1}{2} \frac{m_{\rm imp}(M_{\rm pl} + m)}{(m_{\rm imp} + M_{\rm pl} + m)^2} v_{\rm imp}^2, \quad (20)$$

which scales similarly with impactor properties as $\eta$. This prescription accounts for variation in the impact angle ($\theta$) through the impact parameter $b = \sin(\theta)$. The atmosphere mass loss expressed as fractional atmosphere loss per impactor mass is

$$\frac{m_{\rm atmloss}}{m_{\rm imp}} = \frac{m}{m_{\rm imp}} \times 3.2 \times 10^{-5} \left(\frac{Q}{\rm Jkg^{-1}}\right)^{0.604}. \quad (21)$$

This prescription is a fit to the results of 3D smoothed particle hydrodynamic simulations of atmosphere loss resulting from large impacts, and so should account for both the cratering and non-local atmosphere mass loss.

To use this prescription for our model of the Earth requires significant extrapolation in impactor size and atmosphere mass. Their simulations consider only a single impactor mass and composition (the canonical Moon-forming impactor). The range of atmosphere masses considered is much higher than those considered even in the extremes of our models, from $10^{-2.5} - 10^{-1}$ $M_\oplus$. Furthermore, the atmosphere in their simulations is assumed to be hotter than our atmospheres, which we predict should lead to a higher estimation for the atmosphere mass lost. For a single large impact, this prescription does predict a larger atmosphere mass loss fraction than the S15 prescription although the exact difference depends on the impact properties. Despite these caveats, we consider the effects of using this prescription as an alternative to the S15 prescription in §7.4. To do this we switch from the combined cratering and S15 prescription to the K20 prescription for any impacts with $\eta > 10^9$. This choice is motivated by the fact that this value lies between the maximum $\eta$ considered in Shuvalov (2009) and the approximate minimum $\eta$ considered in Kegerreis et al. (2020).

### 3.3 Full prescription

The values of $\frac{m_{\rm atmloss}}{m_{\rm imp}}$ (including the polar cap limit and large impact induced losses) and $\frac{m_{\rm impacc}}{m_{\rm imp}}$ as a function of impactor size and impact velocity for three atmosphere masses (all with $\mu = 29$) and two impactor densities $\rho_{\rm imp}$ are shown in Figure 1. This figure also shows the locations in parameter space of the simulations performed in Shuvalov (2009), illustrating the regions of parameter space in which these results are constrained by simulations and those where the outcome has been extrapolated. While the fractional atmosphere mass loss is small for both the smallest and largest impactors, the absolute atmosphere mass loss increases monotonically with impactor size. The impactor and target densities used in the Shuvalov (2009) simulations are not quite the same as those used in this work, so these locations are approximate.

Consider first the prescription for mass loss (right panels). The middle row (an Earth-like atmosphere mass), as discussed in Shuvalov (2009), shows that objects with $D \sim 10 - 100$ km are the most efficient at atmosphere removal. Comparing with the top row shows that these most efficient objects become larger as the atmosphere mass increases. The figures also show where the polar cap limit applies, illustrating that this is relevant only for the largest objects (with $D > 100$ km for an Earth-like atmosphere mass), however the impactor size at which it becomes relevant decreases as the atmosphere mass decreases. The effect of non-local atmosphere mass loss caused by the largest impactors is visible only for the most massive atmosphere ($\delta = 8.5 \times 10^{-6}$). Particularly fast large impacts can cause greater atmosphere loss than predicted by the prescriptions discussed above, for example as discussed in Yalinewich & Schlichting (2019). However, the range of impact velocities considered in this work never reaches the extremes considered by that paper, and so their corrections are not included.

Now we consider the prescription for the fractional impactor mass retained. The left panels show that this is constant for impactors larger than $\sim 1$ km. For low density, comet-like, impactors (with density $\rho_{\rm imp} = 0.9$ g cm$^{-3}$) the largest impactors results in no accretion of impactor material by the planet. For higher density, asteroid-like, impactors (with density $\rho_{\rm imp} = 2.8$ g cm$^{-3}$) this is the case only for impact velocities above $\sim 40$ km s$^{-1}$, with a non-zero fraction of the impactor mass accreted for slower impactors. For further discussion of the behaviour in this "airless limit" see §3.4.

### 3.4 The airless limit

Special consideration must be given to the behaviour of the atmosphere in the so called "airless limit" where the atmosphere is negligible in comparison to the impact induced vapour plume. This becomes relevant where the result of impacts is to deplete the atmosphere, but is also important when considering the atmospheric growth (or re-growth) resulting from impacts onto a bare rock. We adopt the approach used in Shuvalov (2009), whereby this limit is defined to apply for impacts with $\eta > 1000$. This occurs for the largest and fastest impactors, and extends to cover a larger portion of the impacting population as the atmosphere mass decreases. In this limit, the fractional impactor mass retained in a collision depends only on the impactor density and velocity, as illustrated in equation 18 and Figure 1. The validity of this approximation is tested through comparison with prescriptions derived for impacts in the absence of atmospheres.

While there are a number of prescriptions available (see, e.g. Thébault & Augereau 2007; Svetsov 2007; Zhu et al. 2019) we compare the Shuvalov (2009) airless limit prescription to that given by Cataldi et al. (2017) as a test of its validity. The prescription presented by Cataldi et al. (2017) (their equations 3, 4 and 5) is based on the experimentally calibrated ejecta model from Housen & Holsapple (2011). A comparison between these prescriptions is shown in Figure 2, from which it can be seen that the qualitative trends in fractional impactor mass retained with impactor size and velocity are the same for both prescriptions. Both prescriptions predict that at low atmosphere masses, the fractional impactor mass retained is independent of impactor mass and depends only on impact velocity. However, the velocity at which the impactor mass being accreted by the target body transitions to zero is lower for the Cataldi et al. (2017) prescription, meaning that either the Cataldi et al. (2017) prescription underestimates the amount of material accreted, or the Shuvalov (2009) prescription overestimates it. The

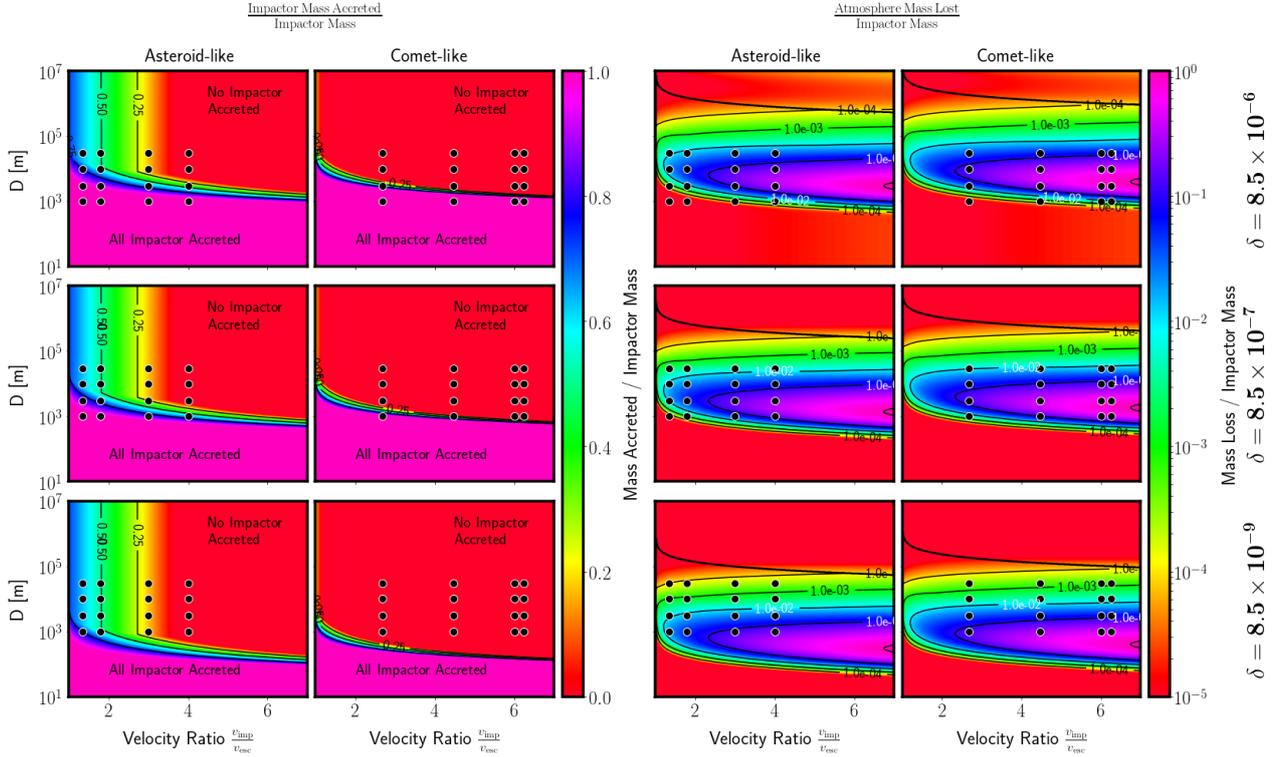

**Figure 1.** Prescription used to calculate the change in atmosphere mass, as a function of impactor size $D$ and the ratio of impact velocity to planet escape velocity $\frac{v_{\rm imp}}{v_{\rm esc}}$. The left columns show the fractional mass accreted $\left(\frac{m_{\rm impacc}}{m_{\rm imp}}\right)$ and the right columns show the fractional atmosphere mass loss $\left(\frac{m_{\rm atmloss}}{m_{\rm imp}}\right)$, including both the polar cap limit (thick black line) and the effect of large impacts (seen most clearly for large, fast, asteroid-like impactors in the most massive atmosphere). The first and third columns show these values for asteroid-like impactors ($\rho_{\rm imp} = 2.8$ g cm$^{-3}$) and the second and fourth columns show comet-like impactors ($\rho_{\rm imp} = 0.9$ g cm$^{-3}$). The locations of the simulations on which these prescriptions are based are shown by filled black circles. These values were calculated assuming an Earth-like planet ($\rho_{\rm pl} = 5.5$ g cm$^{-3}$, $a_{\rm pl} = 1$ au, $M_{\rm pl} = 1$ M$_\oplus$) orbiting a Sun like star. The atmosphere is assumed to have $\mu = 29$, and shown for three different masses, corresponding to 0.01, 1 and 10 times the present day value ($\delta_\oplus = 0.85 \times 10^{-6}$).

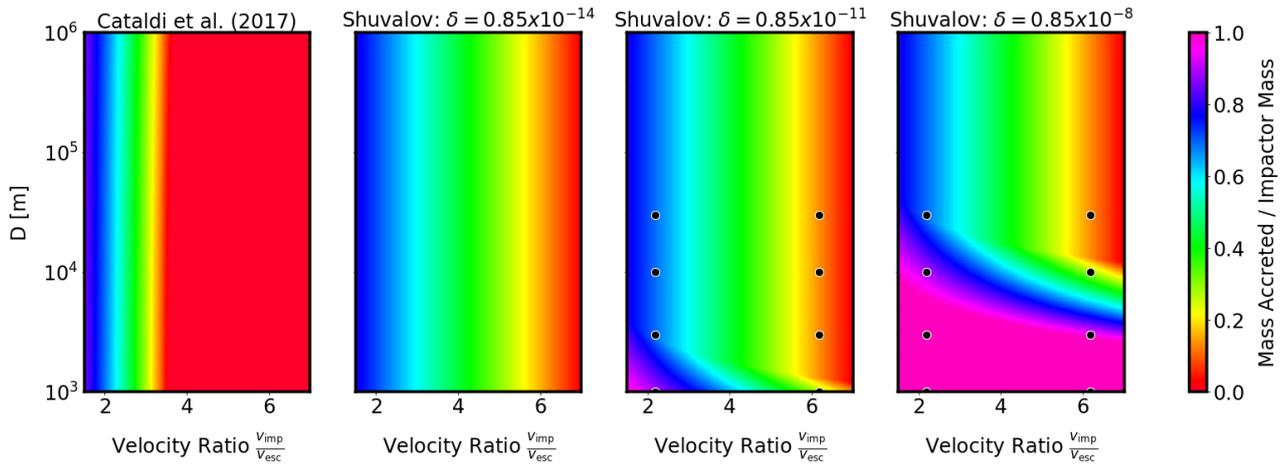

**Figure 2.** Comparison of the predicted fractional mass retained after an impact from the Shuvalov (2009) and Cataldi et al. (2017) expressions, shown as a function of impactor size ($D$) and ratio of impactor velocity to planet escape velocity ($\frac{v_{\rm imp}}{v_{\rm esc}}$) for an Earth-like planet. The Cataldi et al. (2017) prescription assumes no atmosphere, while for the Shuvalov (2009) prescription the results are shown for various different atmosphere masses. Black filled circles show the locations in ($\delta, D, v_{\rm imp}$) parameter space for which simulation or experimental data was obtained.

Housen & Holsapple (2011) models do not distinguish between target and impactor material, and are also based on low-velocity, low mass experimental data. Thus, for the impacts that we consider in this work, where the impactor size and velocity spans a large range, we consider the Shuvalov (2009) prescriptions to be the more relevant, and so adopt them for the remainder of this work. While this potentially misses some detail, it is broadly in agreement with the Cataldi et al. (2017) prescription.

## 4 IMPLEMENTATION AND CODE TESTING

When implementing the code there are choices to be made about the time step and number of bins. These are discussed in §4.1, 4.2 before we outline a set of test simulations for which there is an analytical solution in §4.3 that are used to validate the code and these choices in §4.4.

### 4.1 Time step size choice

Due to the number of impactors considered, it is not computationally feasible to consider individually every single impactor, particularly since the majority of impactors are small and therefore cause negligible atmosphere mass change. However, combining the effects of too many impactors arriving in a single time step could result in a loss of detail in the resulting evolution, or in a total mass of atmosphere loss greater than the atmosphere mass at that time. Therefore we incorporate an adaptive time step. In order to avoid artificially selecting against the largest impactors, which will by definition cause a significant change in the atmosphere, the time step is pre-calculated neglecting the stochastic nature of the impacts. We specify a maximum fraction atmosphere mass change limit (referred to later as the accuracy) $\epsilon = 10^{-4}$ and use this to set the time step length through

$$\Delta t = \epsilon \frac{m}{\dot{m}} \quad \text{where} \quad \dot{m} = \frac{m(t + \delta t) - m(t)}{\delta t}, \quad (22)$$

using equations 1 to 6, modifying the impactor sampling such that it is not stochastic, and therefore the number of impactors of each size is not necessarily a whole number. This calculated value is then bounded from above by a maximum value, set by the time interval of the time dependent impactor properties, and from below by a minimum value of 1 year in order to prevent the code from failing to run.

The choice of the value for the parameter $\epsilon$ is a balance of computation time and accuracy. It must be small enough that the relative change in atmosphere mass remains less than 0.1, when the stochastic nature of the impacts is included, and the atmosphere mass never becomes negative. This is tested in §4.4. We choose initially a value of $\epsilon = 10^{-4}$ as this keeps the relative deviation between the atmosphere masses calculated by the numerical code and the analytic solution $< 0.01\%$ for all but the lowest atmosphere masses.

### 4.2 Bin number choice

The number of impactor size bins to use is determined such that the values of $f_v$ and $t_0$ calculated analytically (through the integrals given in equations 9 − 11 of Wyatt et al. 2019) agree with those calculated using the code, still without including the stochastic nature of the impacts. The resulting convergence of $f_v$, for a range of impactor compositions and velocities, is well fit by a power law

$$\frac{f_v - f_{v,\text{analytic}}}{f_{v,\text{analytic}}} \approx 7.5 \, N_{\text{bins}}^{-2}, \quad (23)$$

with the convergence of $t_0$ following a similar relation. We therefore choose a value of $N_{\text{bins}} = 500$, to achieve convergence to better than 1 part in $10^4$, giving a fractional bin spacing of $\frac{D_{i+1} - D_i}{D_i} \approx 0.03$.

The number of velocity bins is chosen to balance computational cost with capturing the full range and detail of the distribution of impact velocities. These bins are logarithmically spaced between the minimum possible impact velocity (the escape velocity of the planet, 11.2 km s$^{-1}$) and the maximum impact velocity (which depends on the population being considered). We use $N_{\text{vel}} = 50$ and a maximum impact velocity of 85 km s$^{-1}$, which gives a factor of $\frac{v_{j+1} - v_j}{v_j} \approx 0.04$ difference between subsequent bins. For further discussion of the impact velocity distributions, see §5.1.

### 4.3 Test simulations and the analytic solution

In order to test the various features of the code, we construct four test impactor populations that result in a range of atmospheric outcomes. For all tests we consider an Earth-like planet and initial atmosphere ($m_0 = 0.85 \times 10^{-6}$ M$_\oplus$, $\mu = 29$). We consider different impactor compositions, each characterised in terms of a bulk density ($\rho_k$), and volatile fraction ($x_v$). We also assign mean molecular weights ($\mu_k$) to each of the different impactor populations, based on realistic values for the Solar system bodies our test populations most closely resemble. For a full discussion of these values, see §5.4. We assume a total mass contained in the impactor population of $M_{\text{tot}} = 0.01$ M$_\oplus$, and consider a single size distribution, with $D_{\text{min}} = 1$ m, $D_{\text{max}} = 1000$ km and $\alpha = 3.5$. The typical relative velocity of each impactor population is given by $v_{\text{rel}} = \xi v_{\text{pl}}$ which is then converted into an impact velocity (accounting for gravitational focusing)

$$v_{\text{imp}}(\xi) = \sqrt{v_{\text{esc}}^2 + (\xi v_{\text{pl}})^2}. \quad (24)$$

These test impactor properties, labelled by the outcome they have on the atmosphere, are shown in Table 1. For all populations, the value of $f_{v\,j,k}(t)$ is constant in time, and non-zero only in the bin corresponding to $v_{\text{imp}}$.

For each test population, we calculate, using equations 9 − 11 in Wyatt et al. (2019), the values of $f_v$, the ratio of atmosphere growth to loss, and $t_0 = \frac{m_0}{\dot{m}^-}$, the time it would take the atmosphere to be completely depleted in the absence of any volatile delivery. These values combined with the initial atmosphere mass $m_0$, allow the analytic solution to the atmosphere mass as a function of time derived in Wyatt et al. (2019) to be calculated as

$$\frac{m}{m_0} = \left[1 + \left(\frac{\alpha - 1}{3}\right)(f_v - 1)\frac{t}{t_0}\right]^{\left(\frac{3}{\alpha - 1}\right)}. \quad (25)$$

The calculated values of $f_v$ are shown also in Table 1 (for the $\alpha = 3.5$, $D_{\text{max}} = 1000$ km size distribution).

### 4.4 Comparison to the analytic model

We first test the simplest implementation of the numerical code, which calculates the atmosphere mass gain and loss using only the cratering prescriptions given in equations 17 and 13, ignoring the polar cap limit given by equation 16 and the non-local atmosphere mass loss due to large impacts given by equation 19 as these effects are not included in the analytic solution. Furthermore, the stochastic nature of the impacts is ignored (meaning that it is possible for a fraction of an impactor to arrive in a given time step), and the planet mass and atmospheric composition are forced to remain constant in time. In all tests we use 500 size bins, 50 velocity bins and an accuracy of $\epsilon = 10^{-4}$, with minimum and maximum time step sizes of 1 and $10^4$ years respectively. The resulting atmosphere mass is shown as dashed lines in Figure 3a. The general trends in growth or loss for each test population are in agreement with the predictions made using Figure 3 in Wyatt et al. (2019).

**Table 1.** The densities, volatile fractions, mean molecular weights and relative velocities of the four impactor types used in the numerical code to specify the distribution of impactor properties used in the code tests.

| Outcome | Density $\rho_{\rm imp}$ [g cm$^{-3}$] | Volatile Fraction $x_{\rm v}$ | Median Molecular Weight $\mu$ | Relative Velocity Ratio $\xi = \frac{v_{\rm rel}}{v_{\rm pl}}$ | Mass Gain / Loss Ratio $f_{\rm v}$ |
|---|---|---|---|---|---|
| Depletion | 3.5 | $10^{-4}$ | 15 | 0.5 | 0.027 |
| Stalled Depletion | 2.8 | 0.02 | 26 | 1.0 | 0.70 |
| No Change | 2.8 | 0.02 | 26 | 0.9 | 1.03 |
| Growth | 2.8 | 0.02 | 26 | 0.3 | 12.8 |

The deviation of the atmosphere mass calculated assuming the Shuvalov (2009) prescription from the analytic solution is shown in Figure 3b. It is largest for the cases where the atmosphere depletes but is significant even for the "Growth" population. This arises from the fact that the values of $f_{\rm v}$ and $t_0$ are dependent on atmosphere mass, and so change over time. This results in a final atmosphere mass for the "Growth" population that is lower than predicted by the analytic solution. Furthermore, this plot reveals an interesting prediction for the "No Change" and "Stalled Depletion" populations, with the atmosphere mass stalling at a constant value. In the case of the "No Change" population the value of $f_{\rm v}$ is initially very close to one, and so the rates of atmosphere mass gain and mass loss are roughly equal resulting in very little change in the atmosphere mass. For the "Stalled Depletion" population, this phenomenon can be better understood by considering the change in the rate of atmosphere mass gain to mass loss ($f_{\rm v}$) as a function of time. For this impactor population, the value of $f_{\rm v}$ is initially less than one, predicting atmosphere loss, but as the atmosphere mass decreases $f_{\rm v}$ increases until it reaches one and the atmosphere mass loss and mass gain rates are equal resulting in stalled atmosphere mass loss. This phenomenon of an equilibrium solution is is discussed in more detail in Schlichting et al. (2015) and Wyatt et al. (2019). We also show the fractional atmosphere change as a function of time for these tests, in Figure 4. The adaptive time step choice keeps this less than the chosen limit until the atmosphere mass becomes very small (at which point the lower time step limit is reached).

We also consider the inclusion of effects not accounted for by the analytic solution, namely the stochastic nature of impacts, the time evolution of the planet parameters (mass, $M_{\rm pl}$, and mean molecular weight, $\mu$), and the polar cap limit and non-local large impact induced mass loss. Stochasticity is relevant primarily for the largest bodies, since the total numbers of objects in the largest bins are typically less than the number of time steps taken by the code. The atmosphere evolution resulting from including these effects, for the four test impactor populations is shown in Figure 3a as solid lines. The general trends in atmosphere mass over time are similar, but including the stochasticity makes the evolution less smooth with time. This can be seen also in the relative deviation of the code results from the analytic solution shown in Figure 3b. The stochastic effects dominate, with the inclusion of the large impact effects predictably resulting in lower atmosphere masses than when neglecting them, through slower atmospheric growth or more rapid atmospheric loss, depending on the regime the impactors are in. The difference due to time evolution of the planet properties is in general small, because the change in planet mass is small and the atmosphere composition (via $\mu$) is only present in the prescription through $\eta$. The details of the atmosphere evolution resulting from these further effects is not captured by the analytic solution alone, and so motivates the development of this numerical model. The fractional atmosphere change is shown in Figure 4, and can be seen to vary by up to three orders of magnitude between subsequent single time steps, both above and below the specified limit. In all runs it remains less than 100 % unless the atmosphere mass becomes very small (at which point the lower time step limit is reached), and so we do not find it necessary to decrease the chosen value of the accuracy.

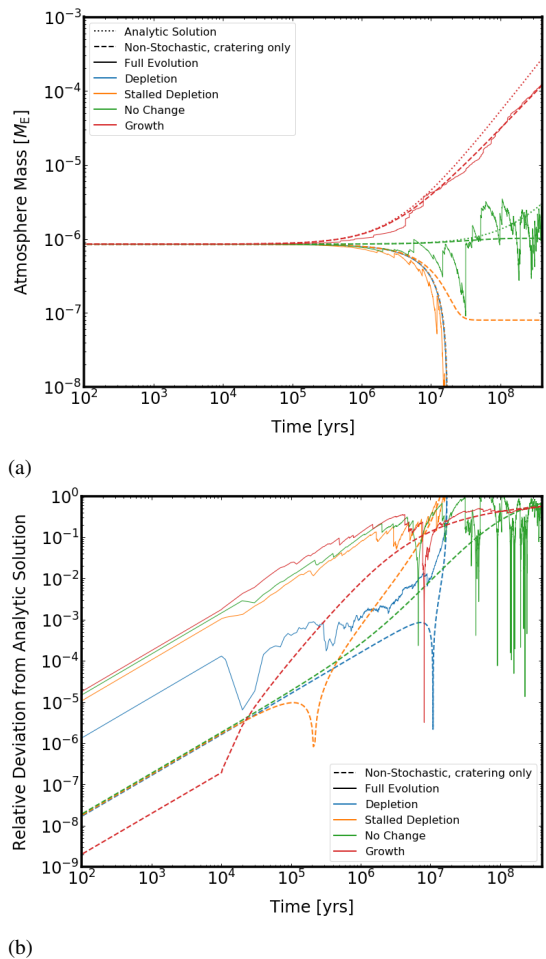

(a)

(b)

**Figure 3.** Comparison of the atmosphere mass calculated using the analytical solution (shown by dotted lines) to the results of the cratering-only implementation of the code (dashed lines) and the full implementation including stochasticity, time evolution of the planet properties and large impacts (solid lines). Panel (a) shows atmosphere mass as a function of time, and (b) shows the relative deviation of the atmosphere mass calculated by the numerical code from the analytical solution. Results are shown for the four test impactor populations described in Table 1, assuming in all cases an $\alpha = 3.5$ size distribution, $D_{\rm max} = 1000$ km and $M_{\rm tot} = 0.01$ M$_\oplus$ for the impactors, and an initially Earth-like atmosphere.

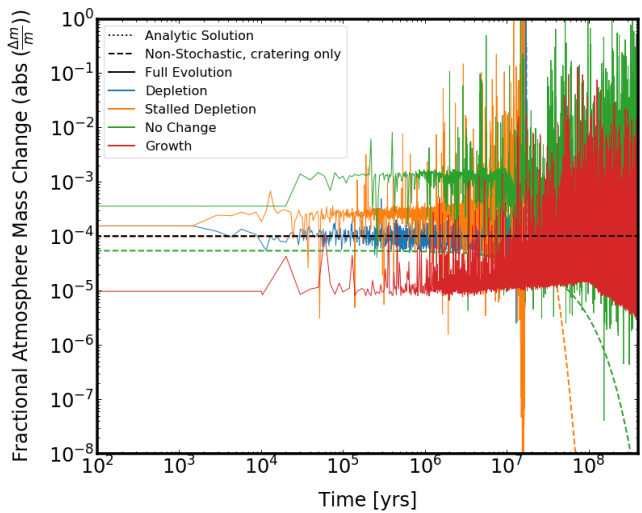

**Figure 4.** The fractional atmosphere mass change as a function of time, for the four impactor populations considered in Figure 3. This limit is constant when the stochastic nature of impacts is ignored, but varies significantly between time steps when the stochasticity is included. The chosen value for the limit ($10^{-4}$) is shown as a black dashed line for reference.

## 5 APPLICATION TO EARTH'S ATMOSPHERE

In order to accurately constrain the evolutionary history of Earth's atmosphere, it is necessary to know the number of objects of different sizes that impact the Earth as a function of time, and the velocities they arrive with as well as the distribution of compositions that they possess. We consider the effect of three distinct populations of impactors: the planetesimals left-over from the terrestrial planet region after terrestrial planet formation (left-over planetesimals), comets destabilised from the trans-Neptunian disk and asteroids ejected from the asteroid belt by the giant planet migration. In the following sections we discuss the choices we make regarding the properties of these populations, and the observational evidence for doing so.

### 5.1 Impact velocities and probabilities

For all impactor populations we make use of the results from the following N-body dynamical simulations to calculate the flux of impactors in each velocity bin.

*Asteroids:* Nesvorný & Morbidelli (2012) performed simulations of the giant planet instability, evolving five giants initially in mutual mean motion resonance, with a disk of left-over planetesimals located beyond the outermost planet. Within these simulations, a series of encounters between Jupiter and Saturn and the fifth giant planet results in discrete, step-like, evolution of the semi-major axes of Jupiter and Saturn, and ejection of the fifth planet. The simulations most successful at recreating the current Solar system architecture are used as the basis for the simulations in Nesvorný et al. (2013), and Nesvorný et al. (2017a) from which we calculate our asteroid impactor fluxes. These simulations are successful in reproducing the orbital distributions of main belt asteroids as well as current impact fluxes. In the case we adopt from Nesvorný et al. (2017a) (their CASE1B), the orbits of the inner planets and 50, 000 asteroids are integrated over the lifetime of the Solar system, using the results from CASE1 in Nesvorný et al. (2013) to determine the orbits of the giant planets during and after the instability (which is assumed to occur early, at $t \approx 5.7$ Myr). The terrestrial planets are assumed to have an initial angular momentum deficit slightly lower than the present value and the initial distribution of asteroids is weighted by a Gaussian distribution in inclination and eccentricity.

*Comets:* To calculate the number of impacts by comets, we use the results from CASE2 in Nesvorný et al. (2017b). These simulations follow the evolution of test particles originating from the trans-Neptunian disk, using artificial force terms to recreate the planetary migration and instability from Nesvorný & Morbidelli (2012). This approximation of the planetary migration is not identical to that assumed for the asteroids, but is similar, and so it is appropriate to use these two cases together here.

For both the asteroid and comet data, time zero in the dynamical simulations does not correspond to our start time (the Moon-forming impact). While the timing of the giant planet instability relative to the Moon-forming impact is not certain, we assume here that it occurred early (Clement et al. 2018, 2019) and thus use data only from times after 50 Myr in the dynamical models.

*Left-over planetesimals:* The simulations of terrestrial planet formation from Walsh et al. (2011) are used as the starting conditions for the simulations of left-over planetesimals performed by Morbidelli et al. (2018). These simulations clone the orbital distributions of left-over planetesimals surviving at 30 and 50 Myr (approximately when the terrestrial planets stabilise), for two different simulations each, and consider two different initial configurations for the terrestrial planets (their current orbits, and circular coplanar orbits). The orbits of the planets and 2000 left-over planetesimals in each of these eight configurations are then integrated for 500 Myr. We consider seven of these configurations (with the one not used randomly excluded), allowing us to account for the uncertainties in the Nice and Grand Tack models. For the nominal case, we consider one configuration, which we call here case 1 which corresponds to one of the cases in which the left-over planetesimals are cloned at 30 Myr, with the terrestrial planets on their current orbits.

The simulations described above give us the orbital elements (semi-major axis, eccentricity and inclination) of each particle in the Earth-crossing region, as well as the orbital elements of the Earth at that time. The Earth was not included in the simulations concerning the comets, so we interpolate the values from the asteroid data for our calculations. For each particle the number of impacts in each velocity bin for each impactor type at each time ($R_k(t) \times f_{v,j,k}(t)$ in §2.1.1) is calculated using the method described in Wyatt et al. (2010). This method represents each particle as a population of particles with random mean longitude, argument of pericentre and longitude of ascending node and does likewise for the Earth, a total of at least $N_{\text{part}} = 10^5$ particles are chosen from the overlapping region of these two populations. In the cases where the Earth is assumed to be on an orbit with zero eccentricity and inclination, the torus representing the Earth is artificially set to have a width of 3 Hill radii[3]. The collision rates and relative velocities between each pair of closest neighbours are calculated and summed (using the relevant weighting) over the orbit to give the probability distribution as a function of relative velocity. The calculated probability distributions (i.e. the probability of a particle on a given orbit colliding with the Earth at a particular velocity) have irregular shapes, often with multiple peaks, that cannot be properly represented by a single average value. This is the motivation behind the inclusion in the numerical code presented in §2 of a distribution from which the impact velocities of the impactors are drawn.

Since the number of particles in the Earth-crossing region is smaller at late times we combine time-steps such that there are always at least $N_{\text{min}} = 50$ particles contributing to the velocity distribution. This choice reflects a balance between retaining the variation in impact velocities resulting from different dynamical histories and avoiding velocity distributions that are inaccurate due to sampling only a small number of particles. The number of comets in the Earth-crossing region declines much more rapidly than the numbers of asteroids and left-over planetesimals, due to ejection through close encounters with Jupiter. As a consequence of this, we are forced combine the velocity distributions for all the comets present after 50 Myr into a single distribution. To avoid the unrealistic scenario of a constant (extremely low) impact rate onto the Earth by comets for the entire time period covered by our simulation, we impose an artificial exponential decay for the impactor flux rate ($R_{\text{comet}}(t)$). This decay is assumed to have a half life of 10 Myr, based on an approximate fit to the decay in total impact probability in time for the comet population. While this is a simplification, we find that the final results are insensitive to the precise impact times compared to the other effects considered.

The distributions calculated using the above method are shown as a mass accretion rate ($\frac{R_k(t)}{\int R_k(t)dt} \times f_{v,j,k}(t) \times M_{\text{tot } k}$, using the total mass estimates described in §5.3) in Figure 5. This illustrates the irregular shape of the distribution at each time step. The results are shown for each of the three impactor populations, with the nominal case for the left-over planetesimals (referred to as case 1 later) shown. The left-over planetesimals are in general slower, and more heavily skewed towards the slowest collisions (at escape velocity), while the asteroid and comet populations both contribute impactors with velocities up to 6 times the escape velocity. The mass accreted from the left-over planetesimal population is significantly higher than the other two populations. The effect of combining multiple time steps to have a minimum of 50 particles contributing to each distribution is most pronounced at late times, when the original simulations contained fewer particles.

---
[3] Increased to 5 Hill radii if 3 gives zero probability

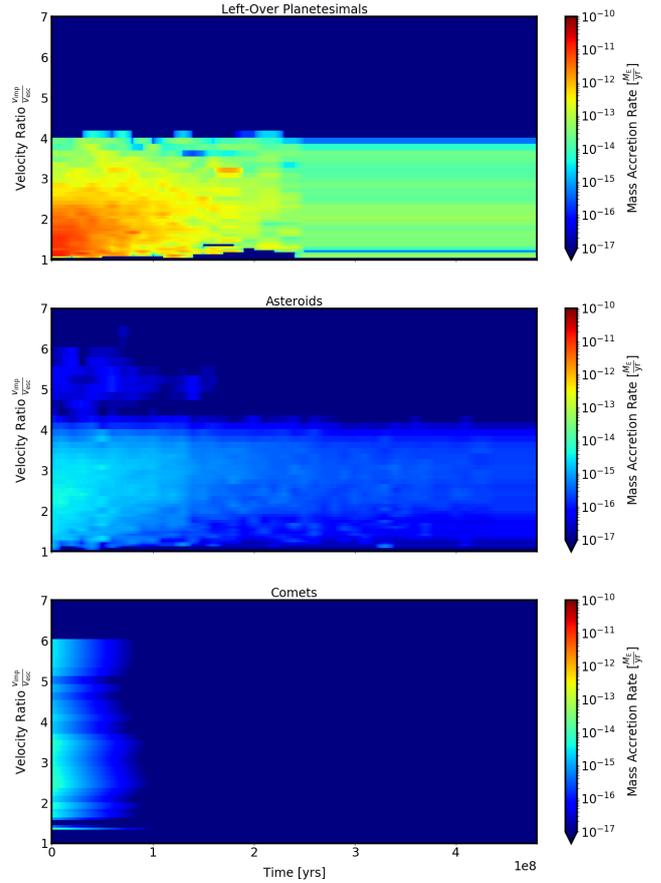

**Figure 5.** An illustration of the distribution of mass accretion rate in time and velocity ($\frac{R_k(t)}{\int R_k(t)dt} f_{vj,k}(t) M_{\text{tot } k}$) for each of the three impactor populations: left-over planetesimals (top), asteroids (middle) and comets (bottom). This presentation allows the total mass accreted from three populations as a function of time or velocity to be visualised through summing over the other axis. These plots shows the distribution after multiple time steps have been combined (if necessary). For comets, the total number of particles used to calculate the distribution is small, meaning that all the velocity distributions are combined. We therefore impose an artificial exponential decay with a half life of 10 Myr.

The calculated values of $\frac{R_k(t)}{\int R_k(t)dt} M_{\text{tot } k}$ and $f_{v,j,k}$ for the different impactor populations are shown in figure 6. We label the left-over planetesimal cases 1 to 7. There can be seen a spike in impact probability around 270 Myr for two cases (3 and 4), due to left-over planetesimals trapped in mean motion resonances with the Earth. These bodies are phase-protected from collisions with the Earth, but they give a non-zero collision probability in our code because resonant protection is not taken into account. After some time these bodies leave the resonance by chaotic diffusion, then quickly disappear under the effects of planetary encounters. The effect of these spikes can be seen in the median atmosphere evolution profiles shown in §6.4.1, but they do not have a significant effect on the final atmosphere masses when compared to the other cases.

### 5.2 Size distribution

The distribution of impactor sizes and compositions at the time of the impacts cannot directly be measured, and so must be inferred from observational signatures. We assume a size distribution that is the same as the present day main belt asteroids, taken from Bottke et al. (2005), for the left-over planetesimal and asteroid populations. This is a simplification, but is supported by observations that the impactor size distribution inferred from the lunar craters matches that expected from the main belt asteroids (Strom et al. 2005). For the comets we assume a shallower size distribution similar to that for the primordial trans-Neptunian disk from Nesvorný et al. (2018). The upper limit of our size distribution is chosen to be $D_{\max} = 1000$ km, approximately equal to the size of the largest object in the asteroid belt. We set the lower limit to be 1 m, and objects between this size and the minimum size given in Bottke et al. (2005) (1 km) or Nesvorný et al. (2018) (100 m) are assumed to have a collisional size frequency distribution that follows an $\alpha = 3.5$ power law (Dohnanyi 1969). These size frequency distributions ($f_{N,i}$) are shown in Figure 7. The largest objects contribute the majority of the total mass in the distribution derived from the main belt asteroid SFD, while the comet mass is dominated by $\sim 100$ km bodies.

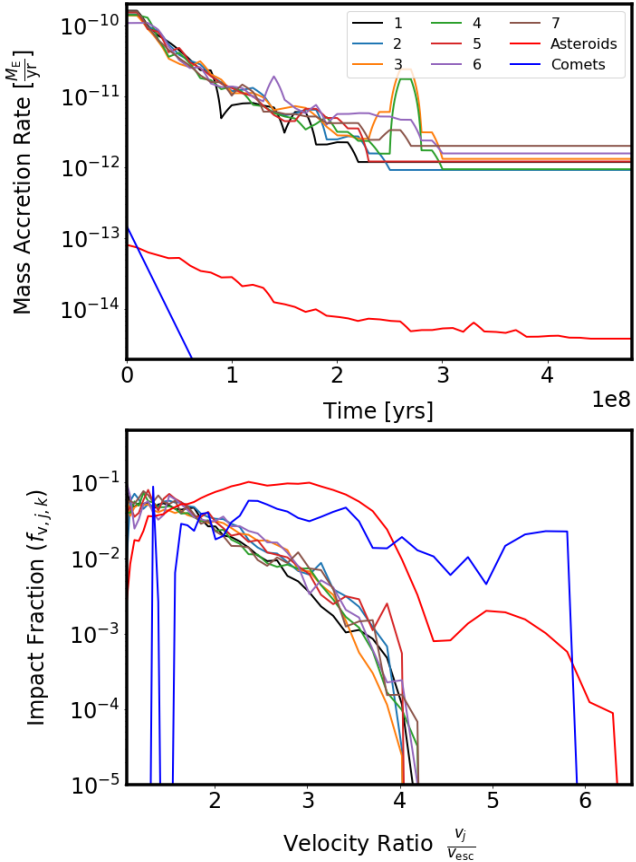

**Figure 6.** The mass accretion rate ($\frac{R_k(t)}{\int R_k(t) \mathrm{d}t} M_{\mathrm{tot}\,k}$) and impact fraction as a function of velocity ($f_{v,j,k}$) calculated for the different impactor populations are shown in the top and bottom plots respectively.

The lower size limit of $D_{\min} = 1$ m is chosen to balance the computational costs of including increasing large numbers of increasingly small impactors with the need for accuracy in our results. Small impactors typically have a negligible influence on the atmosphere as they remove very little atmosphere mass and contain only a fraction of the volatile mass delivered by the entire impactor population. For fixed impactor and planet properties, equation 12 shows that the erosional efficiency depends on atmosphere mass as $\eta \propto m^{-1}$ and thus as the atmosphere mass decreases the value of $\eta$ corresponding to the smallest impactor increases. We find that the lower size limit does not affect the predicted atmosphere evolution provided that the minimum value of the erosional efficiency is $\eta(D_{\min}) < 0.1$. For the range of impactor and planet properties considered in this work this limit is reached only at an atmosphere mass of approximately $10^{-13}$ M$_\oplus$, which is significantly smaller than even the most depleted atmospheres that typically result in our simulations.

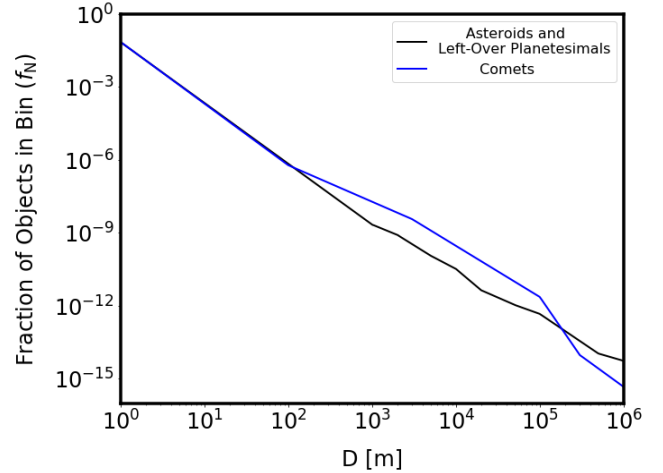

**Figure 7.** The number of objects in each size bin, shown as a black line for the asteroids and left-over planetesimals (based on the asteroid belt size frequency distribution) as a black line, and for the comets (based on the distribution from Nesvorný et al. (2018)) as a blue line.

### 5.3 Total impacting mass

The total number of impactors sampled by the code of each impactor population over all velocity and size bins is normalised by the total mass that *impacts* the planet. This is not the same as the total mass *accreted* by the planet as this depends on the atmosphere mass and composition, which we cannot predict in advance. The total mass that impacts the planet over the course of the simulation ($M_{\mathrm{tot},k}$) is given by

$$M_{\mathrm{tot},k} \equiv \int \sum_i^{N_{\mathrm{size}}} \sum_j^{N_{\mathrm{vel}}} R_k(t) f_{N,i,k}(t) f_{v,j,k}(t) \frac{\pi}{6} \rho_k D_i^3 \, \mathrm{d}t. \quad (26)$$

The choices for these masses are made using different approaches for each of the three populations.

*Asteroids:* For the asteroid population, we make use of the result from Nesvorný et al. (2017a) that 177 asteroids with size $D > 10$ km are expected to impact the Earth over the full 4.5Gyr span of their simulations. Using the exponential fit given by their equation 1, 76 of these impacts are predicted to occur during the 500 Myr time span starting at 50 Myr that we consider. Assuming a weighted average density of the two asteroid types (considered in §5.4) this corresponds to a total mass of asteroids of $M_{\mathrm{tot, ast}} = 3.94 \times 10^{-6}$ M$_\oplus$, which we adopt as our nominal value.

*Comets:* For comets, we normalise the population using the capture fraction of the Jupiter Trojans from the simulations of Nesvorný et al. (2013), which is $f_{\mathrm{capt}} = 5 \times 10^{-7}$ (Nesvorný et al. 2018). The calculation of the intrinsic collision probability described in §5.1 gives that the total probability per comet initially in the disk (after the assumed time of the Moon-forming impact at 50 Myr) is $P_{\rm tot,c} = 1.08 \times 10^{-7}$. Comparing these two numbers, we expect the mass in our comet population to be a fraction $n = \frac{P_{\rm tot,c}}{f_{\rm capt}}$ of the total mass estimated for the Trojans, $M_{\rm Trojans} = (0.3 \pm 0.19) \times 10^{-10}$ $M_\odot$. This gives $M_{\rm tot,c} = 2.2 \times 10^{-6}$ $M_\oplus$ which we take as our nominal value.

*Left-over planetesimals:* The population of left-over planetesimals has no present day population that can be used to normalise it, and so we must take a different approach. We instead normalise this population using the Late Veneer mass (the total mass accreted by Earth since core formation ended, discussed in §1) as an observational constraint. It is implied by isotopic constraints that most of this mass came from left-over planetesimals rather than asteroids and comets (Morbidelli et al. 2018). The normalisation of the left-over planetesimals can therefore not be done directly, as the amount of material accreted necessarily depends on the evolutionary history of the atmosphere, which in turn depends on the total mass of impactors. We therefore use an iterative approach. To calculate the correct normalisation mass, i.e. the total mass contained in the objects that *impact* the Earth) that will result in accretion of a mass approximately equal to the Late Veneer mass, the mass accreted by Earth since core formation (Dauphas 2003), we make a first guess of $M_{\rm tot,\,plan} = 0.01$ $M_\oplus$, then use the code to calculate what mass is accreted onto the planet. We then update the estimate of $M_{\rm tot}$, and repeat this process, until the accreted mass (averaged over several runs to account for stochastic effects) is in agreement with the Late Veneer mass, $(0.005 \pm 0.002)$ $M_\oplus$ (Warren et al. 1999; Walker 2009). In all cases we find a total mass estimate of $M_{\rm tot,plan} = 0.0075$ $M_\oplus$ produces an acceptable range of total accreted masses.

For the comets and asteroids, we considered the uncertainty in their total masses introduced by uncertainties in the values used to calculate them, but this was found to be very small in comparison to the total mass of left-over planetesimals. Instead, we investigate increasing the total mass for these two populations by a factor of ten in §6.2 and §6.1 when these populations are considered in isolation.

### 5.4 Composition

Within the numerical code, composition is defined by a bulk density, volatile fraction, and mean molecular weight. The volatile fraction refers specifically to the mass fraction of the impactor that is outgassed into the atmosphere after the impact. We do not include water in this, on the grounds that it is expected to be in the liquid phase on the surface of planets in the habitable zone (Zahnle et al. 2007). This outgassed mass fraction has an associated $\mu_k$, that depends on the ratios of the different species present. We adopt for each impactor population an approximate representative range for the density and volatile fraction, and assume a linear relation ($x_{\rm v} = A\rho_{\rm imp} + B$) between these two parameters in order to avoid adding a further free parameter. For each population, we use the extrema and centre of these ranges to construct three potential compositions: "wet" (lower density, more volatile rich), "nominal" and "dry" (higher density, more volatile poor). The value of $\mu$ for the mass outgassed by each impactor population is approximated from the molecular abundances in the literature, and assumed to be constant. These parameters are summarised in Table 2 and Figure 8, and their choice justified below.

*Asteroids:* For the asteroid population, we split the population into two sub-populations, which broadly correspond to ordinary chondrites and carbonaceous chondrites in terms of their composition. We use the S-type and ordinary chondrite labels interchangeably (and do the same for C(+B)-type and carbonaceous chondrite), since it allows us to place some necessary constraints on the properties of our impactors. This is imprecise, but until there are better observational constraints on the composition of small Solar system bodies it is necessary to combine data from meteorites and asteroids. In order to avoid confusion, the asteroid types and chondrite labels are used together only in this section, and elsewhere we use only the chondrite labels.

- The C-type asteroid/carbonaceous chondrite-like sub-population is relatively volatile rich. These objects have the bulk of their outgassed volatile content in carbon dioxide, with lesser amounts of hydrogen, carbon monoxide, hydrogen sulfide and sulfates; as well as trace methane, nitrogen and ammonia (Schaefer & Fegley 2010; Sephton 2002). This results in a relatively high $\mu \sim 39$. We adopt a density range spanning $1.5 - 2.5$ g cm$^{-3}$, and a volatile range spanning $0.01 - 0.2$, giving the values of $m$ and $c$ in Table 2 (Carry 2012).
- The S-type asteroid/ordinary chondrite-like sub-population is comparatively volatile poor. These objects have the bulk of their outgassed volatile content in carbon monoxide and hydrogen, with lesser amounts of carbon dioxide, methane, hydrogen sulfide, nitrogen and ammonia, and trace amounts of sulfur dioxide (Schaefer & Fegley 2010). This results in a estimate of $\mu \sim 13$. We assume the density spans $2.5 - 4$ g cm$^{-3}$, and the volatile range spans $0.001 - 0.01$ (Carry 2012).

We distinguish these two populations using their initial semi-major axes ($a_0$), which allows us to calculate two separate distributions of $R_k(t)$ and $f_{{\rm v},j,k}(t)$ for these populations, in all cases assuming the same total masses of C- and S-types summed over the entire belt. To investigate the relative importance of this effect we consider three different cases. First, a flat distribution, with no dependence on $a_0$, splitting the total mass such that $\sim 11.7$ % of the total mass is in S-type asteroids and the rest is in C(+B) types (DeMeo 2014). Second, a highly simplified extrapolation of the taxonomic distribution presented in Table 6 of DeMeo (2014). Considering S-type to represent one population, and combined C and B-types to represent the other, we calculate the mass ratio of S : (B+C) for the inner ($1.6 - 2.5$ au), middle ($2.5 - 2.9$ au) and outer ($2.9 - 4.0$ au) zones of the asteroid belt. Third, an extreme distribution, where we consider all objects with $a_0 < a_{\rm lim}$ to be S-type (ordinary chondrite-like) and all objects with $a_0 > a_{\rm lim}$ assumed to be C-type (carbonaceous chondrite-like). $a_{\rm lim} = 2.5$ au is calculated such that the total mass in each sub-population is consistent with the second case. These are illustrated in the upper plot in Figure 8. The second case is treated as the nominal case when other effects are being investigated.

*Comets:* For the comet population, we assume that the majority of the volatiles are in the form of carbon monoxide, and carbon dioxide, with a small fraction in molecules such as methane, hydrogen sulfide, ammonia, ethane, methanol, formaldehyde, and hydrogen cyanide (Mumma & Charnley 2011; Rubin et al. 2019). This allows us to estimate the value of $\mu \sim 38$ for the volatiles they deliver. In the interests of not overly complicating the prescription we adopt, and noting from Table 3 in Mumma & Charnley (2011) that the range in abundance of a single species is typically at least an order of magnitude, we adopt lower and upper bounds on the volatile fraction of $0.05 < x_{\rm v} < 0.35$, with a most likely value of $x_{\rm v} = 0.2$. This gives a range of densities spanning $0.6 − 1.2$ g cm$^{-3}$.

**Table 2.** The densities and volatile fractions of the four impactor populations used in the numerical code are shown here, both as the parameters describing a linear trend and as nominal values. The estimated value for $\mu$ for the volatiles released is also shown.

| Impactor Type | Minimum $\rho_{\rm imp}$ [g cm$^{-3}$] | Maximum $\rho_{\rm imp}$ [g cm$^{-3}$] | $A$ | $B$ | Nominal $\rho_{\rm imp}$ [g cm$^{-3}$] | Nominal $x_{\rm v}$ | $\mu$ |
|---|---|---|---|---|---|---|---|
| Comet | 0.6 | 1.2 | −0.5 | 0.65 | 0.9 | 0.2 | 38 |
| Carbonaceous / C type | 1.5 | 2.5 | −0.19 | 0.485 | 2.0 | 0.105 | 39 |
| Ordinary / S type | 2.5 | 4.0 | −0.006 | 0.025 | 3.4 | 0.005 | 13 |
| Enstatite | 3.0 | 4.5 | −0.0003 | 0.0014 | 3.5 | 0.00035 | 15 |

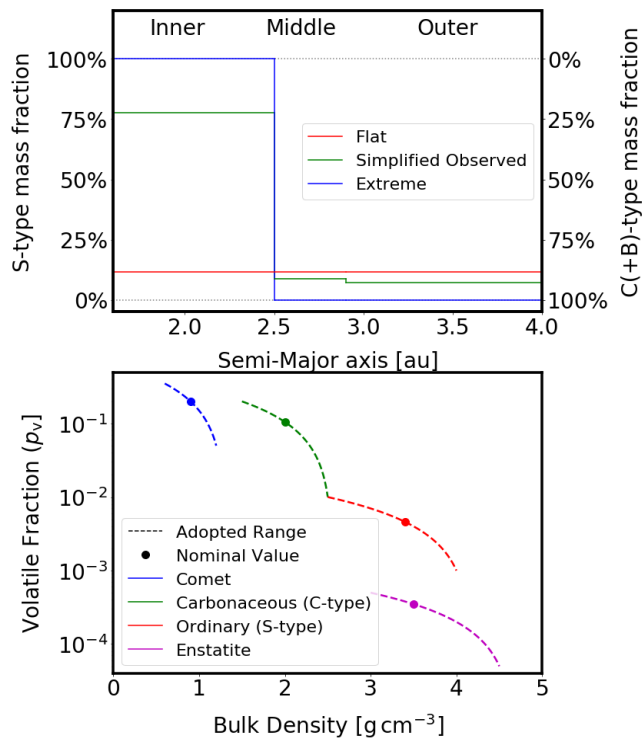

**Figure 8.** Top: the S-type fraction as a function of initial semi-major axis for the three different approaches used to describe the asteroid-like impactor population. The nominal case, a simplified version of the DeMeo (2014) distribution, is shown in green. Bottom: the range (shown by a dashed line) and nominal values (filled circle) of volatile fraction and bulk density of the different impactor populations.

*Left-over planetesimals:* For the population of left-over planetesimals, we assume a very volatile poor enstatite chondrite-like composition. The majority of the material outgassed by this material is in hydrogen and CO, with smaller amounts in carbon dioxide, nitrogen, methane, hydrogen sulfide and ammonia, with trace amounts of sulfur dioxide and other gases (Schaefer & Fegley 2010). This gives an estimate of $\mu \sim 15$. We assume the density spans $3 − 4.5$ g cm$^{-3}$, and the volatile range spans $5 \times 10^{-5} − 5 \times 10^{-4}$ (Carry 2012).

### 5.5 Initial planet and atmosphere conditions

We start our simulations just after the final (Moon-forming) giant impact, after which accreted material cannot be sequestered into the core, and so is recorded in the mantle as an excess of HSEs. The accreted mass, the Late Veneer, is estimated to have a mass of $(0.005 \pm 0.002)$ M$_\oplus$ (Warren et al. 1999; Walker 2009). We therefore assume that the Earth starts in our simulations with a mass of $0.995$ M$_\oplus$. This allows it to reach roughly the present day value of 1 M$_\oplus$ having accreted the Late Veneer mass. This assumes that all the Late Veneer material is delivered in the span of the simulation, which covers 500 Myr, with the material accreted after this time assumed to be negligible. We consider this to be a reasonable approximation, as this period is considered to be the tail end of a period of heavy bombardment, beginning at the end of the giant impact stage of planet formation (Morbidelli et al. 2018).

The Earth is assumed to have constant bulk density ($\rho_{\rm pl} = 5.5$ g cm$^{-3}$), and to be on its current orbit ($a_{\rm pl} = 1$ au). The mass of the Sun is assumed to be equal to its present day value. We adopt a profile for the Sun's luminosity from the models of Bahcall et al. (2001), which increases from approximately $0.7 − 0.75$ L$_\odot$ over the 500 Myr covered by the dynamical data. The assumption that the Earth's atmosphere is isothermal is a simplification, and indeed an adiabatic atmosphere is possibly a more likely scenario in the time immediately following the Moon-forming impact. However it is a necessary assumption in order to make use of the Shuvalov (2009) prescription. The Schlichting et al. (2015) giant impact prescription has an alternate form for adiabatic atmospheres, which takes the same form as equation 19 but with slightly different coefficients. This suggests that there would be a small but non-zero difference in the predicted atmospheric evolution if we did consider a non-isothermal atmosphere.

The mass and composition of the early Earth's atmosphere is not well constrained by observations, but is thought to be broadly similar to the present day (dominated by nitrogen with smaller amounts of carbon dioxide and water) (Kasting 1993). Nitrogen and argon isotope observations have been used to constrain the surface partial pressure of nitrogen to be pN$_2$ < 1.5 bar before 3 Gya (Marty et al. 2013), however more recent models suggest that pN$_2$ was lower in the past (Lammer et al. 2018). Unlike our current atmosphere, the oxygen content was very low before $\sim 2.5$ Gya (Bekker et al. 2004).

We therefore consider a range of values for the initial atmosphere mass, spanning $m_0 = (0.85 \times 10^{-8} − 0.85 \times 10^{-5})$ M$_\oplus$, corresponding to $0.01 − 10$ times the present value. We also consider a range of atmosphere compositions, from $\mu_0 = 2.3$ (representing a primordial hydrogen dominated atmosphere) to 45 (representing a denser, carbon dioxide dominated atmosphere). The effect of this variation in the initial conditions is discussed in §6.6. The nominal values that we adopt when varying other parameters are $m_0 = 0.85 \times 10^{-6}$ M$_\oplus$ and $\mu = 29$, describing the present day atmosphere.

## 6 RESULTS

We first consider the individual effect of each impactor population, which allows us to investigate the result of changing our assumptions regarding the impactor composition in isolation, and the effect of the uncertainty in the dynamics of the population of left-over planetesimals. We are also able to determine which population contributes most significantly to the atmospheric evolution as well as predict what the effect on the atmosphere would be if we vary significantly the estimates we make for the total mass contained in each impactor population. In all cases, the code was run a total of 100 times. Table 3 summarises the results, showing the median and range of the final change in atmosphere mass, final mean molecular weights and fraction of the atmosphere delivered by each impactor considered. Further results, including the total mass *accreted* by the planet for each population and a discussion of the implications of this for water delivery, are shown in Table 5 in §7.3.

### 6.1 Asteroids

We first consider the effect of asteroid impacts on the evolution of Earth's atmosphere, varying both the assumed composition of the asteroids and the spatial distribution of the C- and S-type asteroids. Three compositions ("wet", "nominal" and "dry") and three spatial distributions of the C- and S-type asteroids ("flat", "observed" and "extreme", as described in §5.4) are considered, as well as the effect of increasing the total mass of the impacting population by a factor of ten, giving a total of ten separate cases, each of which was run 100 times.

The evolution of the atmosphere mass through time is shown in Figure 9a, from which it can clearly be seen that the effect of single stochastically sampled events can result in a wide variety of final atmosphere masses. A relative frequency plot of the final atmosphere masses for each of the ten cases is shown in Figure 9b, which also shows the effect of considering only composition (by combining the results from the three different spatial distributions of the asteroid types to give a total of 300 runs in each case). This presentation is motivated by the observation that the distribution of final atmosphere masses is very similar between the "flat", "observed" and "extreme" asteroid distributions that we consider. This is not surprising, as the asteroid impact rates we calculate appear to show little dependence on the spatial distributions of the asteroid types. Thus, in the following, we will neglect this effect, and consider only the combined results to investigate their dependence on the composition and total mass of the asteroid population.

The average value and range for both the final atmosphere mass and fraction of the atmosphere delivered by the two asteroid types are summarised in Table 3. The "nominal" and "wet" asteroids result most typically in atmospheric growth, the final median atmosphere mass growth is similar for the two populations, resulting in percentage changes of 0.86 and 0.60 % respectively, from the initial atmosphere mass of $0.85 \times 10^{-6}$ M$_\oplus$). We also find that the more volatile rich "wet" population produces a larger range of final values than the "nominal" population. The "dry" asteroids in general result in very slight atmosphere loss of $-0.11$ %, however some runs produce an atmosphere mass percentage increase of up to 104 %, more than doubling the atmosphere mass.

The extremely high final atmosphere masses that occur in some runs are the result of the stochastic arrival of large, slow asteroids that deliver a substantial portion of their mass in volatiles to the atmosphere, leading to final atmosphere masses more than an order of magnitude larger than the initial atmosphere mass. This can be understood by considering the maximum mass of volatiles that can be delivered by a single impactor. Equations 12 to 18 give us the impactor mass accreted, and therefore the volatile mass delivered by a single impact,

$$M_{\text{max v}} = \frac{\pi}{6}\rho_{\text{imp}}D^3 x_{\text{v}}(1 - \chi_{\text{pr}}). \quad (27)$$

For the nominal total population mass we consider there are $\sim 0.019$ asteroids with $D > 500$ km, meaning we would expect impactors larger than this size to occur in a few % of our 100 runs in each case. The largest, slowest impactor possibly sampled by the code (with $D = 1000$ km and $\frac{v_{\text{imp}}}{v_{\text{esc}}} = 1.01$, as can be seen in Figure 6) could deliver $\sim 75$ % of its mass, corresponding to a maximum volatile delivery of $M_{\text{max v}} \approx 16\, m_0$, comparable to the largest increases in atmosphere mass seen in Figure 9a. The impactors become more numerous, and therefore large jumps in atmosphere mass become more common, when the total mass of the impacting population is increased.

In contrast to the behaviour observed for the cometary impactors, the "dry" impactors also produce less of the particularly extreme final atmosphere masses seen for the "wet" and "nominal" results. This results from the decreased volatile fraction of these impactors, meaning that the largest possible single delivery of volatiles is smaller, and so stochastic sampling of the same velocity and size of impactors results in a smaller delivery of volatiles to the atmosphere in comparison to the "wet" and "nominal" cases. Increasing the total impactor mass by a factor of ten, results in greater atmosphere growth, with a median final atmosphere change of 32 %, and a greater number of stochastic outliers, since the larger total mass makes the sampling of the largest impactors more probable.

### 6.2 Comets

We now consider the effect on the Earth's atmosphere caused by the population of comets alone, for which the total atmosphere mass as a function of time is shown in Figure 10a. A relative frequency plot of the final atmosphere masses is shown in Figure 10b. From these plots it can be seen that the "wet" and "nominal" comets result in consistent atmospheric erosion, with median total mass decreases of 2.6 - 2.5 % from the initial atmosphere mass of $0.85 \times 10^{-6}$ M$_\oplus$. This erosion is larger by a factor of approximately 10 in the case where the total mass of impacting material increases by a factor of ten. Counter-intuitively, we find that the "drier" impactors results in general in slightly less atmospheric erosion, and in some cases in stochastic atmospheric growth.

To understand this effect, we can consider again the maximum volatile mass that could be delivered by a single large object. From figure 1 we can see that the impactor mass accreted is largest for the lowest velocity impactors, and so we consider the velocity of our slowest comet as an estimate for the velocity ratio $\frac{v_{\rm imp}}{v_{\rm esc}} \sim 1.3$. While for the nominal composition, total population mass and size distribution we predict that there are $\sim 0.033$ objects with $D > 250$ km (and so we would expect impacts by large comets to occur in $\sim 3$ of our 100 runs), equation 18 predicts that even the slowest comets (with $\frac{v_{\rm imp}}{v_{\rm esc}} = 1.3$, as can be seen from Figure 6) result in zero accreted mass for impactors larger than $\sim 10$ km. Therefore only the smallest cometary impactors contribute material to the atmosphere, and these objects are both so numerous that their arrival is not stochastic in nature and so small that an individual impactor cannot deliver a mass of volatiles comparable to the atmosphere mass, meaning that no stochastic large jumps in atmosphere mass occur. However, in the case of the "dry" comets, their density is high enough that they behave more like the asteroids shown in Figure 1. That is to say that a non-zero fraction of even the largest objects can be accreted if it impacts with a low enough velocity. A 250 km diameter "dry" comet contains $8.2 \times 10^{-8}$ $M_\oplus$ in volatiles, comparable to $\sim 10$ % of the initial atmosphere mass. An impact by such an object, if the sampled impact velocity is low and thus a non-zero fraction of the impactor mass was accreted, can explain the few runs that show stochastic increases in atmosphere mass.

| Impactor | Case | Final Change in Atmosphere Mass | | | Final Impactor Fraction | | |
|---|---|---|---|---|---|---|---|
| | | Median [%] | Minimum [%] | Maximum [%] | Median [%] | Minimum [%] | Maximum [%] |
| Asteroid | Wet | 0.60 | −0.25 | 788.76 | 0.97 | 0.23 | 88.77 |
| Asteroid | Nominal | 0.86 | −0.18 | 149.01 | 1.21 | 0.20 | 59.96 |
| Asteroid | Dry | −0.11 | −0.35 | 103.91 | 0.22 | 0.02 | 51.07 |
| Asteroid | Massive | 32.00 | 2.51 | 710.33 | 26.66 | 5.80 | 87.83 |
| Comet | Wet | −2.60 | −2.73 | −2.48 | 0.74 | 0.73 | 0.74 |
| Comet | Nominal | −2.51 | −2.61 | −2.40 | 0.46 | 0.45 | 0.46 |
| Comet | Dry | −2.49 | −2.67 | 0.46 | 0.23 | 0.15 | 3.17 |
| Comet | Massive | −24.1 | −24.5 | −23.9 | 4.75 | 4.73 | 4.76 |
| Planetesimal | Wet | −50.4 | −79.4 | −16.4 | 94.54 | 92.43 | 97.36 |
| Planetesimal | Nominal | −75.3 | −94.7 | −47.4 | 94.94 | 91.46 | 98.10 |
| Planetesimal | Int-1 | −85.9 | −99.2 | −58.2 | 96.12 | 92.32 | 99.98 |
| Planetesimal | Int-2 | −95.4 | −99.98 | −79.3 | 99.12 | 94.72 | 100.00 |
| Planetesimal | Int-3[1] | −99.0 | −100.0 | −84.3 | 100.00 | 97.41 | — |
| Planetesimal | Int-4[1] | −100.0 | −100.0 | −96.9 | — | 98.67 | — |
| Planetesimal | Int-5[1] | −100.0 | −100.0 | −99.1 | — | 97.17 | — |
| Planetesimal | Dry[1] | −100.0 | −100.0 | −100.0 | — | — | — |
| Planetesimal | 1 | −71.9 | −92.0 | −37.2 | 95.00 | 91.96 | 98.13 |
| Planetesimal | 2 | −96.0 | −99.9 | −80.3 | 99.94 | 97.96 | 100.00 |
| Planetesimal | 3 | −77.7 | −93.8 | −36.3 | 99.47 | 95.96 | 99.998 |
| Planetesimal | 4 | −74.6 | −93.2 | −56.0 | 94.97 | 91.73 | 98.07 |
| Planetesimal | 5 | −95.3 | −99.9 | −78.1 | 99.83 | 97.75 | 100.00 |
| Planetesimal | 6 | −92.6 | −99.5 | −70.1 | 99.81 | 98.52 | 100.00 |
| Planetesimal | 7 | −88.0 | −99.7 | −69.4 | 97.87 | 93.50 | 99.98 |

**Table 3.** The final fractional change in atmosphere mass (relative to the initial mass of $0.85 \times 10^{-6}$ $M_\oplus$) and fraction of the final atmosphere mass delivered by the impactor population (impactor fraction), both shown as median, minimum and maximum values, for the 100 runs of the code for each individual impactor population considered. For the asteroids, the results for the three spatial initial distributions of C- and S-types are shown combined, and the impactor fractions are shown as $(f_{S-type}, f_{C-type})$.
[1] These atmospheres were in some cases completely depleted by the arrival of the "dry" left-over planetesimals, so no meaningful final atmosphere impactor fraction can be given.

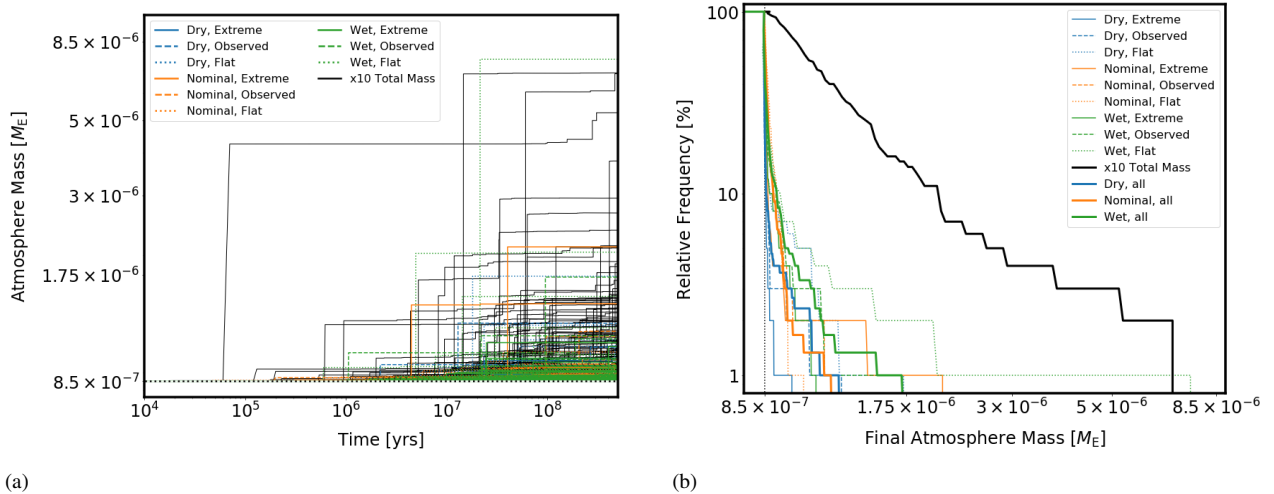

**Figure 9.** The results considering the nominal atmosphere evolved under impacts by the ten different asteroid populations considered. Panel (a) shows the total atmosphere mass as a function of time, and panel (b) shows the corresponding relative frequency (percentage of runs with masses greater than the x-axis value) of the final atmosphere masses. These results are produced by running the numerical code 100 times each, considering three impactor compositions ("wet" "nominal" and "dry", shown by lines of different colours), three potential spatial distributions of the C- and S-type asteroids are also considered ("flat", "observed" and "extreme", shown by different line styles), as well as the effect of increasing the mass of the "nominal" population by a factor of ten.

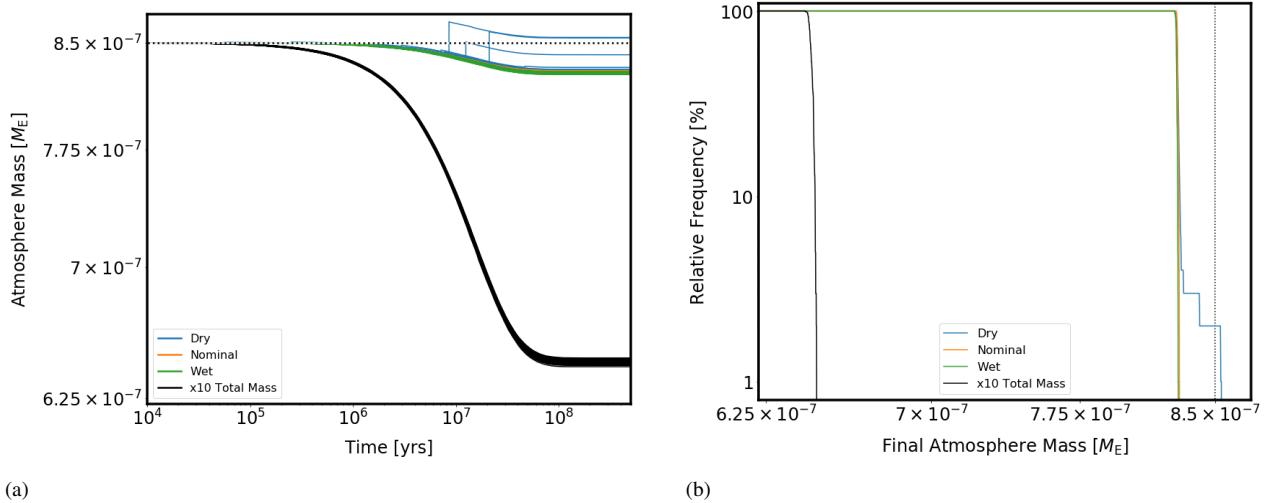

**Figure 10.** The results considering the nominal atmosphere evolved under impacts by the comet population in isolation. Panel (a) shows the total atmosphere mass as a function of time, and panel (b) shows the corresponding relative frequency (percentage of runs with masses greater than the x-axis value) of the final atmosphere masses. These results are produced by running the numerical code 100 times each, considering three impactor compositions ("wet" "nominal" and "dry"), as well as the effect of increasing the mass of the "nominal" population by a factor of ten.

### 6.3 Comparison to previous results

The results presented above for the effect of comets and asteroids on the atmosphere are in contrast to those predicted by the previous study performed by de Niem et al. (2012) (dN12). This work also considered asteroids and comets using a stochastic approach, and so we address the discrepancies briefly here before presenting the results for the left-over planetesimals. For a detailed comparison of our results to those previously published, see §7.5. While both our results and those of dN12 predict atmospheric growth as a result of asteroid impacts, our comets result in atmospheric erosion in contrast to the significant growth predicted by dN12. This difference in behaviour is due to differences in our impact prescriptions and our assumed velocity distribution for the comets. Firstly, the modified implementation of the model from Svetsov (2010) used by dN12 is similar to the Shuvalov (2009) prescription we adopt, but does predict the accretion of a small but non-zero fraction of the largest low density cometary impactor mass. This is in contrast to the Shuvalov (2009) prescription we adopt, which predicts that comets are less efficient at delivering material to the atmospheres. Increasing the density of our comet population results in some fraction of the largest impactor's mass being accreted, switching our results from atmospheric erosion to growth.

Secondly, the velocity distributions adopted by dN12 for their asteroid and comet populations contains a larger number of the slowest impactors. As was discussed in dN12, the slowest impactors can have a disproportionately large effect on atmospheric evolution, since they are in general less efficient at removing atmosphere mass whilst also being more efficient at delivering volatiles. We therefore would expect both more atmosphere erosion and less atmosphere growth for our distribution of velocities, as is observed.

### 6.4 Left-over planetesimals

The material impacting the Earth after the Moon-forming impact is constrained by observations of the Late Veneer to be dominated by an enstatite chondrite-like population of left-over planetesimals and so this population is likely to be the most significant in terms of atmosphere evolution. We consider variation in both the composition and dynamics of the left-over planetesimals.

#### 6.4.1 Dynamics

Considering first the variation in impactor dynamics, Figure 11a shows the results from running the code 100 times assuming the "nominal" composition and comparing the median atmosphere masses at each time for the seven different cases for the dynamics described in §5.1, corresponding to different initial conditions assumed for the N-body simulations. From this plot it can be seen that for each population the atmospheres on average follow a distinct evolutionary track. The distribution of final atmosphere masses for the different populations is shown in Figure 11b, from which the stochasticity induced variation in the final atmosphere masses can be seen. This variation is also apparent from the range of final atmosphere mass changes summarised in Table 3, although it is significantly smaller than that observed for the asteroid-like impactors.

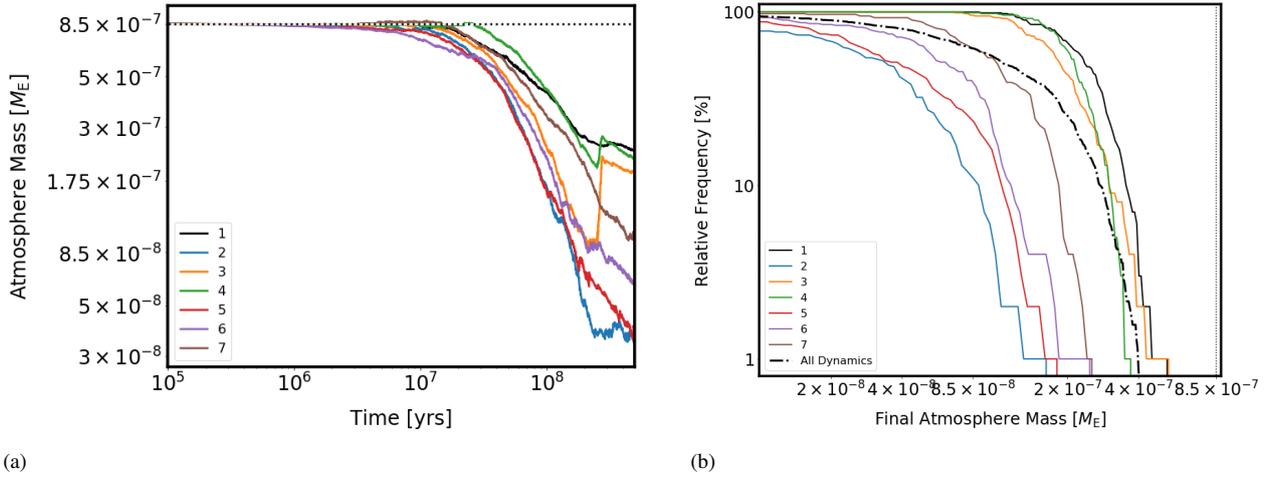

**Figure 11.** The results considering the nominal atmosphere evolved under impacts by the populations of left-over planetesimals with different dynamics. Panel (a) shows the median total atmosphere mass as a function of time, and panel (b) shows the corresponding relative frequency (percentage of runs with masses greater than the x-axis value) of the final atmosphere masses. These results are produced by running the numerical code 100 times each. The seven different cases for the planetesimal dynamics are shown as lines of different colours and all assumed to have the nominal composition.

The general behaviour differs between the different runs, i.e. between the different impactor dynamics, however all seven cases result in atmospheric erosion. The atmosphere mass loss is more pronounced in some runs than in others, with the median final atmosphere mass varying between 4 and 38 % of the initial value for cases 2 and 1 respectively. No atmospheres in these cases are completely stripped, but do end up less than 0.1 % of the initial value in one or two runs for cases 2 and 5. These cases correspond to two different angular momentum deficits with a 30 Myr cloning time in the Walsh et al. (2011) simulations from which the left-over planetesimal orbits are taken. From these results it would seem that the assumptions made about the total angular momentum deficit of the terrestrial planets has a smaller effect on the evolution of the atmosphere mass than the assumptions made about the timescale for terrestrial planet formation and the initial orbits of the terrestrial planets.

The key factor determining the final atmosphere mass is the number of particularly slow impactors sampled by the numerical code, a phenomenon that has been noted previously in de Niem et al. (2012). The seven different distributions of planetesimal impactors as a function of impact velocity shown in Figure 6 look similar, but in fact differ in the number of very slow impactors that they predict. Using the analytical model of Wyatt et al. (2019), for an $\alpha = 3.5$ power law size distribution of impactors with the nominal left-over planetesimal composition, we can predict the value of $f_v$ (the ratio of atmosphere growth to erosion) as a function of impact velocity. As noted in Wyatt et al. (2019), this transition can be very sharp, and for the toy model here we predict that velocities greater than $v_{\rm imp} \approx 1.06 v_{\rm esc}$ should result in atmosphere erosion, while slower velocities should result in growth. The behaviour of our population is more complicated than this toy model, but in general we would expect that sampling from distributions with more of the slowest (atmosphere growing) impactors should result in higher final atmosphere masses. This is demonstrated in Figure 12, which illustrates the dependence of the relative change in median final atmosphere mass for each of the seven populations on the fraction of impactors with velocity below this limit. As predicted, cases with fewer slow impactors end with smaller final atmosphere masses, as the impacts in general tend to be more eroding.

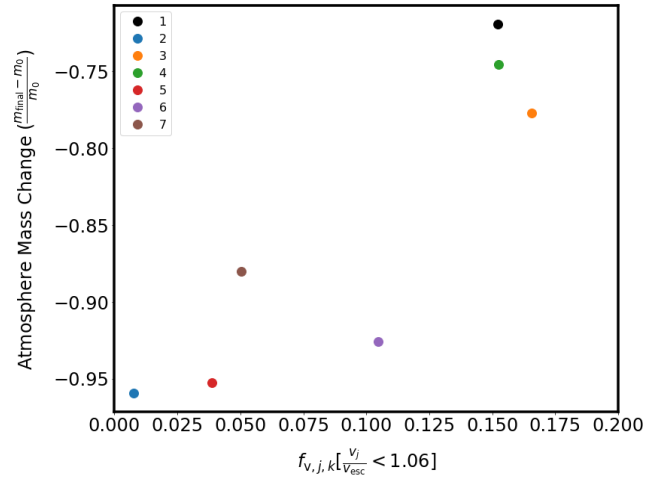

**Figure 12.** The median relative change in final atmosphere mass change as a function of the fraction of impactors with velocities below the value predicted (by a simplified analytic model) for the transition from atmospheric growth to erosion.

The average final composition of these atmospheres, summarised in Table 3, varies slightly depending on the degree of atmospheric erosion. Comparing the different cases, the median final impactor-derived atmosphere fraction varies between 95.0 % for case 1 (which resulted in the least atmospheric erosion) and 99.9 % for case 2 (which resulted in the largest median atmosphere erosion).

| Label | $\rho_{\rm imp}$ [g cm$^{-3}$] | $x_{\rm v}$ |
|---|---|---|
| Wet | 3.0 | 0.0005 |
| Nominal | 3.5 | 0.00035 |
| Int-1 | 3.75 | 0.000275 |
| Int-2 | 4.0 | 0.0002 |
| Int-3 | 4.125 | 0.0001675 |
| Int-4 | 4.25 | 0.000125 |
| Int-5 | 4.375 | 0.0000875 |
| Dry | 4.5 | 0.00005 |
| C-Type | 2.0 | 0.105 |

**Table 4.** The label, bulk density and volatile fraction adopted for the three original and five extra compositions considered for the populations of left-over planetesimals.

*6.4.2 Composition*

We investigate the composition of the left-over planetesimal population in more detail than for the other two impactor populations due to the sensitivity of the final atmosphere mass to the assumed composition, which can be seen in Figure 13a. This is achieved using nine different populations: "wet", "nominal", "dry", five values intermediate between "nominal" and "dry" (which are assumed to follow the same linear relation between density and volatile content described in §5.4) and an extreme case similar to C-type asteroids. These properties are summarised in Table 4.

Figure 13a shows the median atmosphere mass as a function of time, which can be seen to depend strongly on the assumed impactor composition. The distribution of final atmosphere masses for the different populations is shown in Figure 13b, and the median value and range of the final change in atmosphere mass are given in Table 3. There is an obvious lack of large stochastic increases in atmosphere mass, such as those seen for the asteroid and "dry" comet populations. This can again be explained by considering the maximum mass of volatiles that could be delivered by a single impactor. In this case, there are $\sim 20$ objects with $D > 500$ km, and so we would expect "large" impacts to occur in all runs of the code. With our nominal parameters, the maximum mass of volatiles delivered by the slowest, largest impactor possible (with $\frac{v_{\rm imp}}{v_{\rm esc}} = 1.01$ and $D = 1000$ km) is $M_{\rm max\ v} = 0.099\ m_0$. This also explains the observed decrease in absolute magnitude of variation in the final atmosphere mass about the median value as the volatile content of the impactors is reduced, because the largest, slowest impactors will contain comparatively less volatiles, and so the stochastic effect of single impacts will be smaller in this case.

As the volatile content of the impactors is decreased the median final atmosphere mass decreases. Two transitions in the distribution of final atmosphere masses occur. Firstly, between the "Int-2" and "Int-3" compositions, some fraction of the simulations result in loss of the majority of the atmosphere mass.[4] Secondly, between the "Int-5" and "dry" populations no atmospheres survive the length of the simulation (and in the case of "Int-5" this represents only one out of the 100 runs). This total stripping occurs most rapidly for the "dry" population, with a median time for depletion that is approximately 64 Myr.

---

[4] In order to prevent an unreasonably long computation time, the code is halted once the atmosphere mass reaches $10^{-15}$ M$_\oplus$. This results in an under-estimation of the total solid mass accreted in these runs, but test runs of individual cases do not suggest that the atmosphere is capable of recovering from this level of atmosphere loss and thus this approach does not affect our estimation of the minimum volatile content required for atmosphere survival

These transitions are explained by the very low volatile content assumed for these impactors, meaning that on average the population cannot deliver sufficient mass in volatiles to balance the atmosphere mass they remove upon impact. The minimum volatile content for all the atmospheres to survive (albeit at typically 5 % of the original atmosphere mass) is that of case Int-2 which has $x_{\rm v} = 0.02$ %. Cases Int-3, Int-4 and Int-5 result in total atmosphere loss in some runs, but not all, implying that a volatile content greater than $x_{\rm v} = 0.01675$ % (corresponding to case Int-3) is required to guarantee that the atmosphere is not entirely stripped. As discussed in §5.2, our choice of minimum impactor size (1 m) is valid at atmosphere masses greater than $\sim 10^{-13}$ M$_\oplus$. When considering depleted atmospheres in detail the volatile delivery by smaller impactors should be accounted for, but we do not investigate this in detail here.

Increasing the volatile content of the impactors to a level comparable to the C-type asteroids results in significant atmosphere growth, with a median final atmosphere mass of approximately $300 - 480\ m_0$. If the Late Veneer mass was predominantly delivered by this kind of impactor, as opposed to drier enstatite chondrite-like material, we would therefore predict a final atmosphere mass orders of magnitude greater than the present day value.

The average final composition of these atmospheres, summarised in Table 3, is also determined by the impactor composition. In general, as the volatile content of the impactor population decreases, and so the final atmosphere mass decreases, the proportion of the final atmosphere mass that has been delivered by the population of left-over planetesimals increases. We also observe a smaller range of final proportions for the "drier" impactors, as a result of the smaller range of final atmosphere masses noted above.

### 6.5 Representative evolution

In order to investigate the effects of stochasticity on the atmosphere evolution, we run the code with the nominal values for all parameters described in the previous section for a total of 500 iterations, now including all three impactor populations. The resulting atmospheric evolution for all runs is shown in Figure 14a, which shows the atmosphere mass as a function of time. A relative frequency plot of the final atmosphere masses for these 500 runs is shown in Figure 14b. From this it can be seen that the majority of runs follow a very similar profile, with the atmosphere mass increasing very slightly to a maximum value by around $1 - 50$ Myr, then decreasing steadily over the remainder of the simulation. The median final atmosphere mass ($\pm 95$ % confidence intervals) is $(0.238^{+0.543}_{-0.150} \times 10^{-6})$ M$_\oplus$. This represents a loss of $\sim 72$ % of the atmosphere mass.

The median atmosphere mass as a function of time for these runs is shown in Figure 15, along with the median contribution to the atmosphere from the different impactor populations (assuming the "nominal" values for their compositions) over time. From this it can be seen that the "typical" results all end up with left-over planetesimal dominated atmospheres. The median ($\pm 95$ % confidence intervals) compositions of the final atmospheres are $(93.8^{+2.7}_{-3.6})$ % left-over planetesimal, $(0.685^{+0.370}_{-0.481})$ % C-type

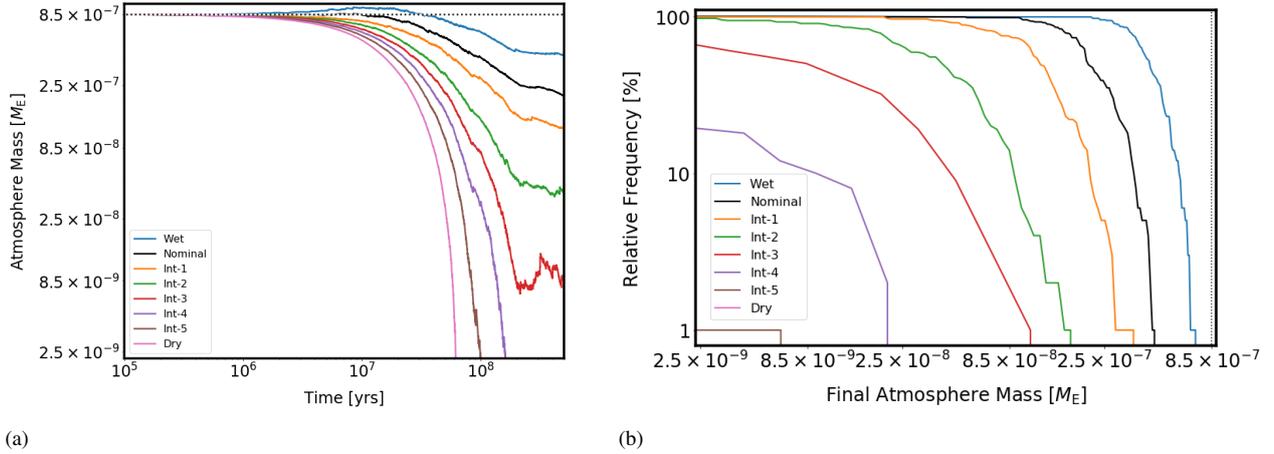

**Figure 13.** The results considering the nominal atmosphere evolved under impacts by the populations of left-over planetesimals with different assumed compositions. Panel (a) shows the median total atmosphere mass as a function of time, and panel (b) shows the corresponding relative frequency (percentage of runs with masses greater than the x-axis value) of the final atmosphere masses. These results are produced by running the numerical code 100 times each. The eight different left-over planetesimal compositions considered are shown as lines of different colours, and all assumed to have the nominal (case 1) dynamics.

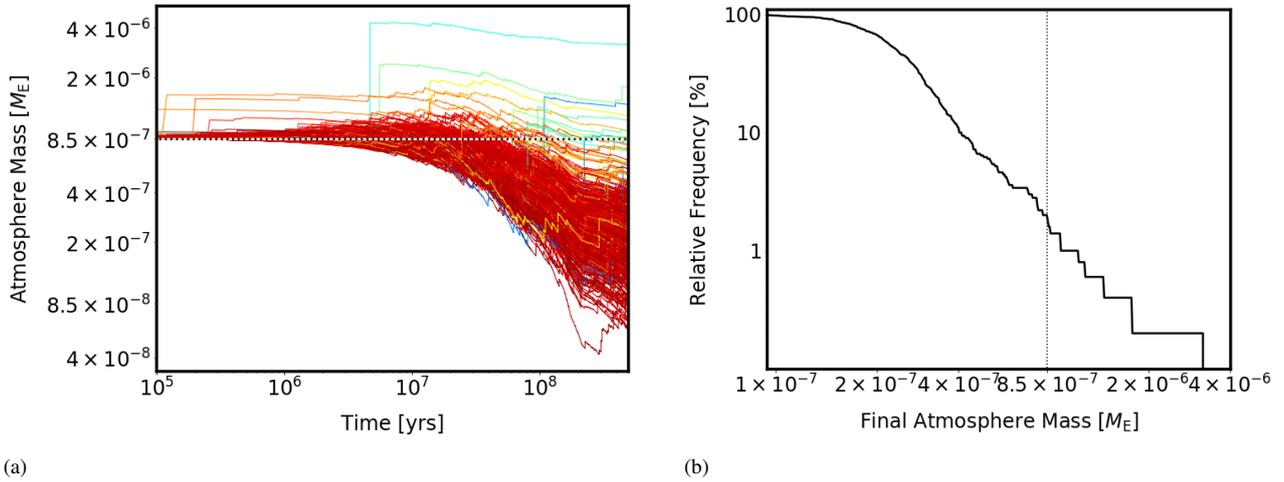

**Figure 14.** The results produced for 500 iterations of the numerical code considering the evolution of an Earth like atmosphere and planet, impacted by populations of comets, asteroids and left-over planetesimals. Panel (a) shows the total atmosphere mass as a function of time, and panel (b) shows the corresponding relative frequency (percentage of runs with masses greater than the x-axis value) of the final atmosphere masses. The colour of the line in panel (a) reflects the composition of the final atmosphere, with bluer lines representing more asteroid dominated atmospheres and redder lines more planetesimal dominated. The impactor populations' dynamics, size distribution, total mass and composition are set by the nominal values of the free parameters.

asteroidal, $(0.0692^{+0.0.0412}_{-0.0.0364})$ % cometary, and $(0.004^{+0.323}_{-0.003})$ % S-type asteroidal. From this we can conclude that approximately three quarters of the original primary atmosphere is lost, and replaced with mainly volatiles delivered by the population of left-over planetesimals, with small, variable contributions from C- and S-type asteroids, and a small but less variable contribution from comets. The cometary contribution is consistent with the conclusion inferred from the atmospheric noble gas budget that cometary material made up < 0.5 % of the late accretion mass (Marty et al. 2016; Schlichting & Mukhopadhyay 2018).

It has been proposed that comets may be an important source of noble gasses in the Earth's atmosphere. Xenon is of particular interest, since atmospheric xenon is depleted relative to both chondritic xenon and atmospheric krypton, and is also more enriched in its heavy isotopes than any potential source. This combination of constraints is impossible to achieve through models of atmospheric escape, giving rise to the so-called "Xenon Paradox". One potential explanation to this paradox is the delivery of ∼ 22 % of the Earth's atmospheric Xe by cometary material (Marty et al. 2017). While only ∼ 0.07 % of our final atmosphere mass is typically made up of material delivered by comets, this is not necessarily in conflict with the above statement. Comets are disproportionately enriched in Xe relative to both carbon and hydrogen compared to other Solar system bodies (Halliday 2013). It is therefore possible that a small amount of cometary material could contribute significantly to the atmosphere's overall Xe inventory. We leave a detailed investigation of the elemental abundances and isotope ratios of the Earth's atmosphere for a future study.

The fractional contribution of each source (primary, left-over planetesimals, both asteroid types and comets) to the final atmosphere mass for each of the 500 runs is shown in Figure

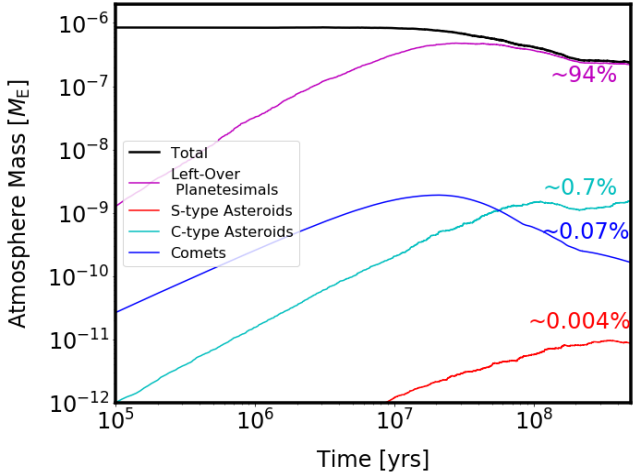

**Figure 15.** The median total atmospheric mass as a function of time for the average evolution of the 500 code runs using the nominal impactor parameters. The median proportion of the atmosphere mass that has been delivered by each of the different impactor populations at each time is shown also.

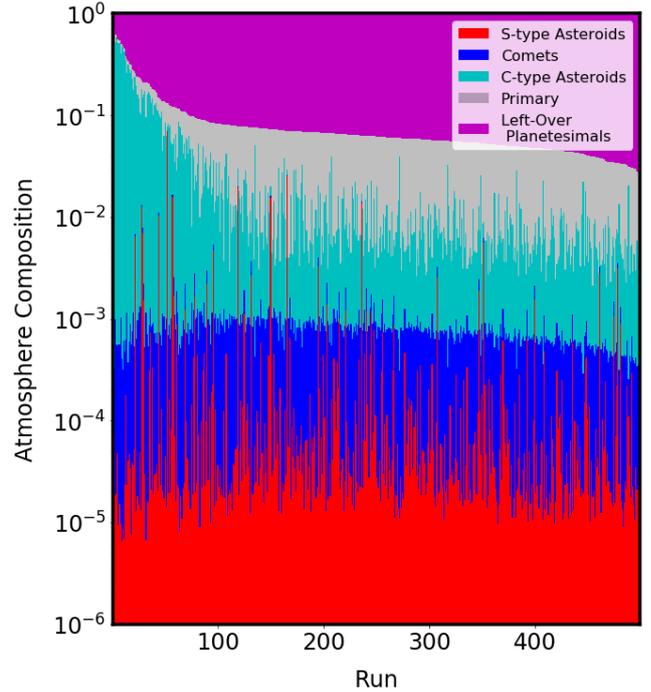

**Figure 16.** The fractional source of the final atmosphere resulting from the 500 runs shown in Figure 14a, in order of increasing planetesimal portion, shown on a log scale. The majority of runs result in atmospheres that are dominated by the volatiles delivered by the population of left-over planetesimals, with a smaller portion due to residual primary atmosphere, followed by C-type asteroids, comets and finally S-type asteroids. A small fraction of the runs show unusually large final atmosphere masses, and are dominated by the volatiles delivered by single large C-type asteroid impacts, or in two cases, an S-type asteroid impact.

16. The majority of runs result in the planetesimal dominated atmospheres discussed above. However, the effects of random, rare impacts sampled stochastically by the code cannot be ignored, and are illustrated here by the 1.4 % ≈ 7/500 of runs that produce unusually high final atmosphere masses. These correspond to the bars in figure 16 where the final atmosphere is dominated by volatiles delivered by C-type asteroids. These atmospheres are heavily influenced by single impacts by large asteroids, with final atmosphere masses of up to $3.2 \times 10^{-6}$ $M_\oplus$. These large asteroids are slow, sampled from the top left hand corner of the asteroid-like plots in Figure 1, and result in modest atmosphere mass loss but contribute a significant fraction of their volatiles to the atmosphere. As discussed in §6.1, a large, slow C-type atmosphere can deliver a mass of up to approximately $10^{-6}$ $M_\oplus$ of volatiles in a single impact. It is therefore possible for a single impactor to cause these deviations from the typical final atmosphere mass. These impactors would have a mass of ∼ $2 \times 10^{-5}$ $M_\oplus$, larger than the total mass we estimate the combined population of C-type asteroids to have by a factor of ∼ 6, meaning that these kinds of impactors should be sampled by the code very rarely. This behaviour is not likely to be representative of the evolution of our own atmosphere, as isotopic analysis suggests that the Late Veneer was delivered by enstatite chondrite-like material (our population of left-over planetesimals). However there has been recent work suggesting that it is possible that the Late Veneer could consist of either entirely non-carbonaceous material (as we assume), or a combination of carbonaceous and non-carbonaceous material (Hopp et al. 2020).

We also track the fractional change in atmosphere mass within each time step, which is kept below ∼ 10 % even when large stochastically sampled impactors arrive, illustrating the successful implementation of the adaptive time step.

### 6.6 Initial conditions

We now consider how the choice of initial atmospheric mass ($m_0$) and initial atmosphere mean molecular weight ($\mu_0$) affects the evolution of the atmosphere. We consider four different initial atmosphere masses, spanning four orders of magnitude from $m_0 = (0.85 \times 10^{-8} - 0.85 \times 10^{-5})$ $M_\oplus$. We also consider four different initial atmosphere compositions, with $\mu_0 = 2.3, 15, 29$ and $45$, giving us sixteen different initial configurations. These compositions roughly correspond to a primordial hydrogen dominated atmosphere, a steam atmosphere comprised of mainly water and some hydrogen, a nitrogen dominated Earth-like atmosphere, and a mainly carbon dioxide atmosphere like that found on Venus. All other parameters for the impactor populations are kept at the nominal values. The median evolution of these initial atmospheres by the four impactor populations described in §5 for these different initial conditions is shown in Figure 17a.

The stochastic effects are more pronounced for the atmospheres that start with smaller masses, as a fraction of the total atmosphere mass can be delivered by a relatively smaller impactor. This results in a greater relative range of final atmosphere masses for these populations, which can be seen in the distributions of final atmosphere masses shown in Figure 17b. The change in atmosphere properties is illustrated in Figure 18, which shows the initial and final locations of all the atmospheres considered, as well as the median final values of each set of initial conditions in atmosphere mass-mean molecular weight space.

From this we can see that in general the initially low mass atmospheres ($m_0 = 0.85 \times 10^{-8}$ $M_\oplus$) undergo growth, tending towards the same final condition regardless of their starting compo-sition, with median final masses between $\sim 6.6 - 9.9 \times 10^{-8}$ M$_\oplus$. The atmospheres that start with a mass 10 % of the present day value undergo relatively minor changes in atmosphere mass, with median final masses between $\sim 7.9 - 9.5 \times 10^{-8}$ M$_\oplus$, but all tend towards a common composition dominated by the material delivered by the population of left-over planetesimals. The atmospheres that start with the present day mass in general deplete, with the primordial ($\mu_0 = 2.3$) case having a median final atmosphere within the range reached by the low initial atmosphere masses. The median final atmosphere mass in this case increases with increasing $\mu_0$, from $7.4 \times 10^{-8}$ M$_\oplus$ to $4.6 \times 10^{-5}$ M$_\oplus$, but still show atmosphere loss. For the largest initial atmospheres (ten times the present atmosphere mass), the final atmosphere mass depends strongly on the composition, with a higher $\mu_0$ resulting in a higher final atmosphere mass ($6.9 \times 10^{-6}$ M$_\oplus$ for $\mu_0 = 45$ compared to $4.1 \times 10^{-6}$ M$_\oplus$ for $\mu_0 = 15$). This effect is most pronounced for the primordial composition, which depletes to a level similar to that reached by the initially less massive atmospheres ($2.3 \times 10^{-7}$ M$_\oplus$). These results suggest that atmospheres with lower $\mu_0$ are easier to remove through impacts, and that the properties of the final atmosphere can be determined entirely by the impactor population if that population is capable of completely replacing the initial atmosphere.

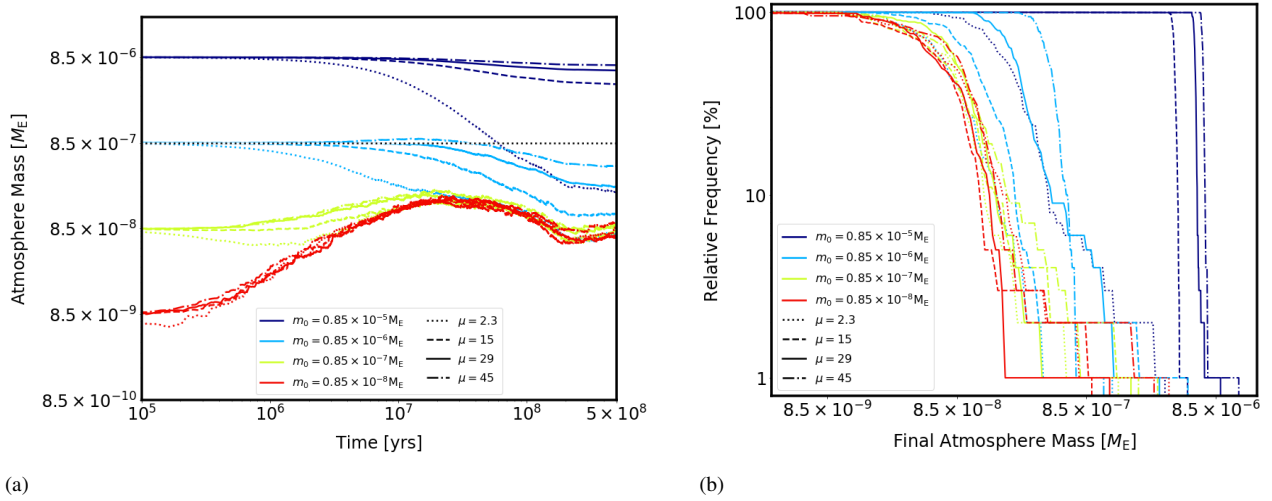

**Figure 17.** The results produced considering the evolution of sixteen different initial atmospheres, run 100 times each, under bombardment by the nominal populations of comets, asteroids and left-over planetesimals. Panel (a) shows the median total atmosphere mass as a function of time, and panel (b) shows the relative frequency (percentage of runs with masses greater than the x-axis value) of the final atmosphere masses. The initial atmosphere mass is indicated by the line colour, with the $\mu_0$ shown by the line style.

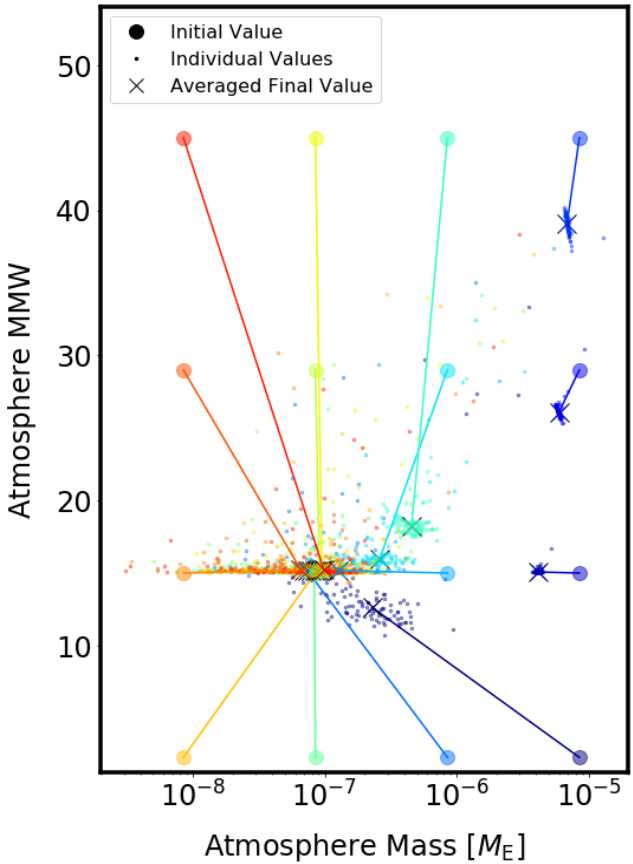

**Figure 18.** The change in the atmosphere mass and mean molecular weight over the course of our simulations for the sixteen different initial atmosphere conditions. The initial conditions are shown as filled circles of different colours. The final values are shown as crosses at the median location, with the final values shown in dots of the same colour.

## 7 DISCUSSION

### 7.1 Variation in the impactor population parameters

The results of §6.2, 6.1 and 6.4 can be used to consider how changing the assumptions made about the populations of comets, asteroids and left-over planetesimals might affect these conclusions. Consider the average evolution of the atmosphere in the representative (all "nominal" parameters) case, which results in approximately half the original atmosphere mass being lost, with a final atmosphere composition dominated by the material delivered by the population of left-over planetesimals. If the population of comets was much more massive, or was "wetter" (lower density and higher volatile content) we would not necessarily expect to see much change in the atmosphere evolution, however if they were "drier", we might expect to see occasional cases where the final atmosphere mass is large and dominated by material delivered by the comets. These atmospheres would be analogous to the rare large final atmosphere masses that we see in the representative case due to the stochastic sampling of a large, slow, asteroid. If the asteroid population was much more massive, we predict that the final atmosphere mass would be higher, with the delivery of asteroid material potentially able to negate the atmospheric loss caused by the nominal population of left-over planetesimals. We would also expect in this case to see the final atmosphere composition dominated by asteroidal material, with the C- and S-type fractions in proportion to their total mass ratio (around 8.6 : 1, see §5.4). In general we would expect a "wetter" population of asteroids to result in a larger final atmosphere mass. This would also result in an increase in the occurrence of stochastic large impacts leading to a wider range of predicted final atmosphere masses. A drier population of left-over planetesimals would be expected to strip the entire atmosphere mass very efficiently, within 75 Myr, while a "wetter" population would be expected to cause less erosion than the "nominal" case, this is discussed in more detail in §6.4. The dynamics of the left-over planetesimals appears to be a less important parameter, however our results suggest that variation in the median final atmosphere mass of $\sim 25\%$ is possible (comparable to variation between a volatile content of $x_v = 0.02\%$ and $0.035\%$).

### 7.2 Atmospheric Convergence

The convergence of the final atmosphere towards a mass and bulk mean molecular weight determined only by the properties of the impactor populations regardless of the initial atmospheric conditions, as shown in §6.6, is an interesting outcome. The stalling of an atmosphere at a particular mass is a phenomena that was predicted first by Schlichting et al. (2015). The behaviour observed in our results can be understood using the formalism of the analytical model of Wyatt et al. (2019). This is because for a range of plausible impactor and atmosphere properties the ratio of atmosphere growth rate to loss ratio ($f_v$, calculated through equation 7) can decrease with increasing atmosphere mass. If it crosses unity at some atmosphere mass, then atmospheres initially above this mass will deplete, while those below it will grow, and both will be expected to stall at $m_{\text{stall}}$, where the rate of growth matches the rate of loss. If instead $f_v$ increases with atmosphere mass, this represents an unstable stalling mass. This behaviour is illustrated in Figure 7 of Wyatt et al. (2019).

The impactor populations we have considered in this work are significantly more complicated in their properties than those of the analytical model, but we can make simplifications in order to investigate the stalling effect. Adopting our nominal impactor compositions and total masses, and considering each population in isolation, assuming an $\alpha = 3.1$ power-law slope for the size distribution of the comet population, and an $\alpha = 3.5$ power-law slope for the remaining three populations, we can calculate $f_v$ (the ratio of atmosphere growth to loss as a function of atmosphere mass) averaged over the distribution of impact velocities given by $f_{v,j,k}$, discussed in §5.1. This is shown for the four impactor populations in Figure 19.

From this we can see that, in agreement with the results presented in §6.2, this analytic model predicts atmospheric loss ($f_v < 1$) as a result of cometary impacts. The results for asteroids are also in line with what we predict, with the dominant

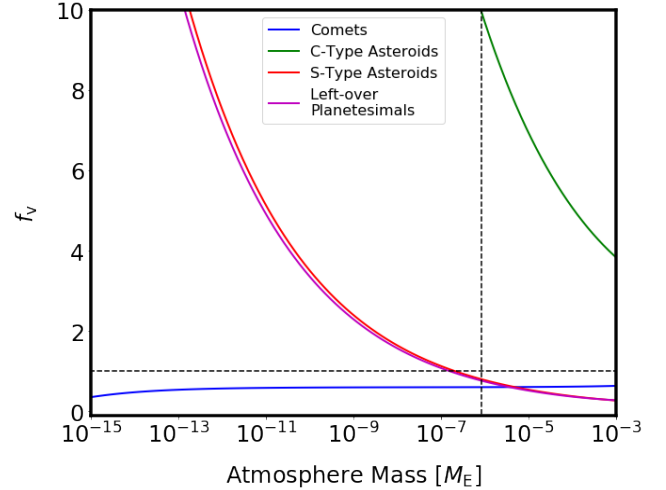

**Figure 19.** The ratio of atmosphere growth to loss as a function of atmosphere mass calculated according to equation 7, averaged over the distribution of impact velocities for each population, and assuming simplified properties representative of each of our four impactor populations. The line $f_v = 1$, where the atmosphere mass should remain constant, is shown by a horizontal dashed black line. The current atmosphere mass of the Earth is shown by a vertical dashed black line.

C-type population predicting atmospheric growth. Impacts by the population of left-over planetesimals are predicted to result in atmospheric loss, as we find using the full numerical model. Due to the significantly higher mass of impacting left-over planetesimals in comparison to the other populations, we would expect the atmospheric evolution and thus the predicted stalling mass to be dominated by the effect of this population. Figure 19 would imply that planetesimal impacts should result in an atmosphere that stalls at a mass of approximately $1.5 \times 10^{-7}$ $M_\oplus$, consistent with what we find for this nominal population using the full numerical code.

The observed convergence might imply that it is not possible to constrain the initial conditions of the Earth's atmosphere in the period just after the Moon-forming impact, before the accretion of the Late Veneer. However, while there may be no signature of the initial atmospheric mass and composition in the final mass and composition it remains to be seen whether more detailed isotopic signatures (for example in the $^{15}$N/$^{14}$N ratios, or in the noble gases) might be capable of distinguishing between different initial scenarios. This relies on data regarding the isotopic signatures of the volatiles contained within the different impactor populations, but is an interesting avenue of research.

While it is not possible to infer precisely the initial conditions for the Earth's atmosphere, the results of §6.6 can be used to place some speculative limits on the initial atmosphere of the Earth. While we might expect the atmosphere to grow slightly after the end of this period of bombardment due to outgassing, it is unlikely that significant atmospheric loss occurred after this time. Therefore we can conclude that an atmosphere larger than 10 times the present mass is unlikely to have been present on Earth after the Moon-forming impact, unless such an atmosphere had a primordial composition, and was therefore easier to remove via impacts.

### 7.3 Water delivery

We can further use the results of §6.2, 6.1 and 6.4 to estimate the approximate amount of water delivered to Earth by each of our impactor populations. Using these results rather than the combined results from §6.5 allows us to investigate how changing the assumptions made regarding the impactor composition and dynamics affects the predictions for the water delivery. This is a simplification, but comparing the nominal single impactor results to the representative case with all populations considered, we find no significant difference in the water mass delivered by each population.

We have excluded water from the volatile inventory that we track, as we assume it will be in liquid form at the atmosphere temperatures we consider and so will not contribute to the atmosphere. However we can estimate the water delivered by combining estimates for the average bulk water content of the different impactor populations with our results for the amount of solid impactor material accreted by the Earth. We adopt approximate water fractions $H_2O$ (wt. %) for the nominal populations of comets, asteroids (weighted average of C- and S-types), and left-over planetesimals of 50 % (Mumma & Charnley 2011), 10 % and 0.001 % respecitvely (Barnes et al. 2016). The water fraction is assumed to be constant between our "wet", "nominal" and "dry" compositions, in order to avoid additional complexity, with the caveat that the masses calculated are rough estimates. Using these we calculate the mass of water delivered by each of the impactor populations considered.

The results are shown in Table 5, from which it can be seen that we predict that for the typical outcomes the largest contributors are the left-over planetesimals, which deliver around 0.2 % of an ocean mass of water ($M_{\rm ocean} = 2.3 \times 10^{-4}\ M_\oplus$). This prediction is roughly constant between different assumptions about the planetesimal properties, but could vary if the water content of the left-over planetesimals is substantially different from our assumed value. If instead the population of left-over planetesimals had a composition more similar to C-type asteroids, they could deliver at least an ocean mass of water to Earth, however as discussed in §6.4.2 this would result in an unrealistically large final atmosphere mass. Higher levels of water delivery (up to 0.04 $M_{\rm ocean}$) are possible in the extreme cases where single large asteroids impact the Earth.

The total water content of the Earth is very uncertain, with estimates of between $0.25 - 4 M_{\rm ocean}$ water potentially present in the mantle (Ahrens 1989; Jambon & Zimmermann 1990; Bolfan-Casanova et al. 2003), and up to $5 M_{\rm ocean}$ water present in the core (Wu et al. 2018). The origin of this water is also not well understood, it may have been accreted during the early stages of planet formation, prior to the Moon-forming impact (Drake & Campins 2006), or it may have been delivered by the Moon-forming impactor (Budde et al. 2019). The fraction of Earth's water delivered by comets is still uncertain, but there is evidence that it is at most 10 % (Dauphas et al. 2000), which is in agreement with our predictions. To exceed this limit would require either a much more massive comet population, or an even drier composition (with a density greater than 1.2 g cm$^{-3}$ and volatile fraction less than $x_{\rm v} = 0.05$, recalling the non-intuitive result of §6.2) or some combination of the two factors.

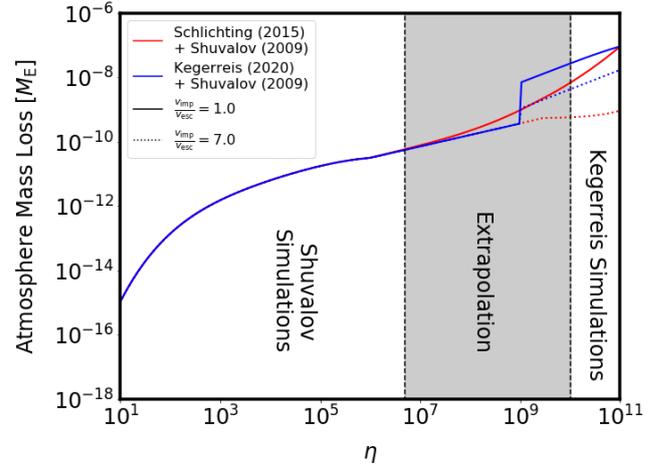

**Figure 20.** The atmosphere mass loss calculated by the combined cratering and S15 prescription we use in the majority of this work, and the alternative combined cratering and K20 prescription (assuming a 45 deg impact angle), for an Earth-like atmosphere and C-type asteroid-like impactors. In order to display the large values of $\eta$ at which these prescriptions diverge, we calculate the predicted atmosphere mass loss for impactors with sizes $D = 1 - 10^7$ m, and two extremes of impact velocity. The range of values for $\eta$ corresponding to the cratering simulations, and the K20 SPH simulations are shown, as is the (shaded) region between the two in which the cratering prescription is extrapolated.

### 7.4 Alternative large impact prescription

To investigate the dependence of our results on the assumptions made about the impact prescriptions we consider replacing the S15 large impact prescription described in §3.2 with the prescription from K20. The atmosphere mass loss predicted by the two prescriptions we consider is shown in Figure 20, from which we can see that the K20 prescription does predict higher mass loss for both the most and least energetic impactors, for a range of impact velocities. The approximate locations covered by the two sets of simulation results are shown, while the region of parameter space in which the results must be extrapolated is highlighted.

The Shuvalov (2009) prescriptions assume $\theta = 45$ deg as the most likely impact angle, while S15 is based on head-on ($\theta = 0$ deg) calculations. Due to the strong dependence of the K20 predicted atmosphere loss on this parameter, we consider both the results for assuming a constant impact angle of 45 deg and sampling the impact angle for each impactor randomly from a uniform distribution. As discussed in §3.2, we predict that the hotter atmospheres used in the simulations on which it is based should result in an overestimation of the atmosphere mass loss when applied to an atmosphere with a temperature of 273 K as assumed in this work. Running our code on the reference atmosphere with this new prescription replacing the S15 prescription we find that it predicts slightly greater atmosphere mass loss. This is illustrated in Figure 21, which shows the relative frequency of the final atmosphere masses obtained in our reference case (as presented in §6.5) compared to the results for the same initial conditions and impactor populations, but implementing the K20 prescription.

In general the distributions of final atmospheres masses are similar, with median percentage changes in the final atmosphere mass of $-72.0$, $-71.2$ and $-76.9$ % for the S15, K20 (45 deg) and K20 (sampled impact angle) prescriptions respectively. Kruskal-Wallis H Tests show that the K20 (45 deg) results are not statistically different from the combined cratering and S15 prescription, (with p-value $p = 0.153$), however the K20 (sampled angle) results are ($p = 2.3 \times 10^{-5}$). These results, using randomly sampled impact angles, result in cases of greater atmospheric loss and also of more extreme large atmosphere masses than the 45 deg case. This is likely due to the non-linear dependence of the atmosphere mass loss on impact angle (see equations 20 and 21). The higher number of low final atmosphere masses in the K20 prescription runs in general agrees with our prediction made in §3.2, that this prescription might overestimate the atmosphere mass loss by the largest impactors due to the higher atmospheric temperatures used in their simulations. However, this variation is smaller than that arising from the variation in impactor dynamics considered in §6.4.1, and so we can consider our conclusions robust to the assumptions we make regarding the large impact prescription.

| Impactor | Case | $H_2O$ (wt. %) (bulk estimate) | Impactor Mass Accreted [$M_\oplus$] | | | Water Delivered [$M_{ocean}$] | | |
|---|---|---|---|---|---|---|---|---|
| | | | Median | Minimum | Maximum | Median | Minimum | Maximum |
| Asteroid | Wet | 10 | $6.27 \times 10^{-8}$ | $1.17 \times 10^{-8}$ | $3.35 \times 10^{-5}$ | $2.72 \times 10^{-5}$ | $5.09 \times 10^{-6}$ | $1.46 \times 10^{-2}$ |
| Asteroid | Nominal | 10 | $1.28 \times 10^{-7}$ | $1.88 \times 10^{-8}$ | $1.21 \times 10^{-5}$ | $5.56 \times 10^{-5}$ | $8.16 \times 10^{-6}$ | $5.27 \times 10^{-3}$ |
| Asteroid | Dry | 10 | $2.18 \times 10^{-7}$ | $2.73 \times 10^{-8}$ | $8.86 \times 10^{-5}$ | $9.46 \times 10^{-5}$ | $1.18 \times 10^{-5}$ | $3.85 \times 10^{-2}$ |
| Asteroid | Massive | 10 | $3.76 \times 10^{-6}$ | $6.67 \times 10^{-7}$ | $6.15 \times 10^{-5}$ | $1.64 \times 10^{-3}$ | $2.90 \times 10^{-4}$ | $2.67 \times 10^{-2}$ |
| Comet | Wet | 50 | $1.77 \times 10^{-8}$ | $1.75 \times 10^{-8}$ | $1.78 \times 10^{-8}$ | $3.85 \times 10^{-5}$ | $3.81 \times 10^{-5}$ | $3.88 \times 10^{-5}$ |
| Comet | Nominal | 50 | $1.92 \times 10^{-8}$ | $1.90 \times 10^{-8}$ | $1.95 \times 10^{-8}$ | $4.18 \times 10^{-5}$ | $4.13 \times 10^{-5}$ | $4.24 \times 10^{-5}$ |
| Comet | Dry | 50 | $3.86 \times 10^{-8}$ | $2.58 \times 10^{-8}$ | $5.49 \times 10^{-7}$ | $8.39 \times 10^{-5}$ | $5.61 \times 10^{-5}$ | $1.19 \times 10^{-3}$ |
| Comet | Massive | 50 | $1.81 \times 10^{-7}$ | $1.80 \times 10^{-7}$ | $1.81 \times 10^{-7}$ | $3.93 \times 10^{-4}$ | $3.91 \times 10^{-4}$ | $3.94 \times 10^{-4}$ |
| Planetesimal | Wet | 0.01 | $4.65 \times 10^{-3}$ | $3.46 \times 10^{-3}$ | $5.71 \times 10^{-3}$ | $2.02 \times 10^{-3}$ | $1.50 \times 10^{-3}$ | $2.48 \times 10^{-3}$ |
| Planetesimal | Nominal | 0.01 | $4.97 \times 10^{-3}$ | $3.45 \times 10^{-3}$ | $6.61 \times 10^{-3}$ | $2.16 \times 10^{-3}$ | $1.50 \times 10^{-3}$ | $2.87 \times 10^{-3}$ |
| Planetesimal | Int-1 | 0.01 | $5.16 \times 10^{-3}$ | $3.38 \times 10^{-3}$ | $6.79 \times 10^{-3}$ | $2.25 \times 10^{-3}$ | $1.47 \times 10^{-3}$ | $2.95 \times 10^{-3}$ |
| Planetesimal | Int-2 | 0.01 | $5.27 \times 10^{-3}$ | $4.15 \times 10^{-3}$ | $7.40 \times 10^{-3}$ | $2.29 \times 10^{-3}$ | $1.81 \times 10^{-3}$ | $3.22 \times 10^{-3}$ |
| Planetesimal | Int-3[1] | 0.01 | $5.18 \times 10^{-3}$ | $3.55 \times 10^{-3}$ | $7.80 \times 10^{-3}$ | $2.25 \times 10^{-3}$ | $1.54 \times 10^{-3}$ | $3.39 \times 10^{-3}$ |
| Planetesimal | Int-4[1] | 0.01 | $4.97 \times 10^{-3}$ | $3.56 \times 10^{-3}$ | $7.43 \times 10^{-3}$ | $2.16 \times 10^{-3}$ | $1.55 \times 10^{-3}$ | $3.23 \times 10^{-3}$ |
| Planetesimal | Int-5[1] | 0.01 | $4.73 \times 10^{-3}$ | $3.55 \times 10^{-3}$ | $7.53 \times 10^{-3}$ | $2.06 \times 10^{-3}$ | $1.55 \times 10^{-3}$ | $3.27 \times 10^{-3}$ |
| Planetesimal | Dry[1] | 0.01 | $4.15 \times 10^{-3}$ | $3.01 \times 10^{-3}$ | $5.85 \times 10^{-3}$ | $1.81 \times 10^{-3}$ | $1.31 \times 10^{-3}$ | $2.54 \times 10^{-3}$ |
| Planetesimal | 1 | 0.01 | $5.08 \times 10^{-3}$ | $3.95 \times 10^{-3}$ | $6.48 \times 10^{-3}$ | $2.21 \times 10^{-3}$ | $1.72 \times 10^{-3}$ | $2.82 \times 10^{-3}$ |
| Planetesimal | 2 | 0.01 | $4.94 \times 10^{-3}$ | $3.87 \times 10^{-3}$ | $6.31 \times 10^{-3}$ | $2.15 \times 10^{-3}$ | $1.68 \times 10^{-3}$ | $2.74 \times 10^{-3}$ |
| Planetesimal | 3 | 0.01 | $4.85 \times 10^{-3}$ | $3.79 \times 10^{-3}$ | $6.44 \times 10^{-3}$ | $2.11 \times 10^{-3}$ | $1.65 \times 10^{-3}$ | $2.80 \times 10^{-3}$ |
| Planetesimal | 4 | 0.01 | $4.95 \times 10^{-3}$ | $3.75 \times 10^{-3}$ | $6.49 \times 10^{-3}$ | $2.15 \times 10^{-3}$ | $1.63 \times 10^{-3}$ | $2.82 \times 10^{-3}$ |
| Planetesimal | 5 | 0.01 | $4.79 \times 10^{-3}$ | $3.23 \times 10^{-3}$ | $6.13 \times 10^{-3}$ | $2.08 \times 10^{-3}$ | $1.41 \times 10^{-3}$ | $2.66 \times 10^{-3}$ |
| Planetesimal | 6 | 0.01 | $4.82 \times 10^{-3}$ | $3.79 \times 10^{-3}$ | $6.44 \times 10^{-3}$ | $2.10 \times 10^{-3}$ | $1.65 \times 10^{-3}$ | $2.80 \times 10^{-3}$ |
| Planetesimal | 7 | 0.01 | $4.89 \times 10^{-3}$ | $3.56 \times 10^{-3}$ | $6.14 \times 10^{-3}$ | $2.13 \times 10^{-3}$ | $1.55 \times 10^{-3}$ | $2.67 \times 10^{-3}$ |
| **ALL** | Reference | – | $5.00 \times 10^{-3}$ | $3.68 \times 10^{-3}$ | $7.46 \times 10^{-3}$ | $2.32 \times 10^{-3}$ | $1.68 \times 10^{-3}$ | $2.11 \times 10^{-2}$ |

**Table 5.** The final change in planet mass and calculated water mass delivered by the impactor population, both shown as median values and ranges, for the 100 runs of the code for each individual impactor population considered. For the asteroids, the results for the three initial distributions of C- and S-types are shown combined, and the impactor fractions are shown as ($f_{S-type}, f_{C-type}$). Note [1] These atmospheres were in some cases completely depleted and so the code halted early, in which case the final solid mass accreted (and therefore water delivered) is underestimated. The final line shows the water mass delivered by all four population in the reference case.

## 7.5 Comparison with previous studies

Our conclusions regarding the potential evolution of Earth's atmosphere can be compared to those of Pham et al. (2011) (P11), de Niem et al. (2012) (dN12) and Wyatt et al. (2019) (W19). Each of these papers adopted different impact prescriptions, and assumed different properties for the impactor populations. Combined with the different methods of calculating the atmospheric evolution over time these result in different predictions for the evolution of Earth's atmosphere during accretion of the Late Veneer. Here we briefly summarise the approach, inputs and results for each paper and discuss how our approach in this paper has led to sometimes significantly different conclusions.

### 7.5.1 *Pham et al. (2011)*

P11 investigated the different evolutionary pathways that the atmospheres of Earth, Venus and Mars may have taken as a result of impacts. The effect of bombardment on the three planets was considered using a simplified prescription for atmosphere mass loss, by dividing impactors into two categories: those that are too small to have any effect on the atmosphere, and those that are massive enough to completely remove the entire atmosphere mass contained in the polar cap ($m_{cap}$). Rather than using the results of N-body simulations to calculate impactor flux, they assume an exponential decay in impacts from $t_0 = 4.6$ Gya to the present day, and assume that a fixed fraction of the impactors have velocities high enough to cause the loss of the polar cap mass. To calculate the mass of volatiles delivered in an impact they assume another parameterisation in terms of the planet mass, volatile fraction and vaporisation factors for each impactor population. They consider only asteroids and comets, and calibrate the impact fluxes using the lunar crater record, estimating a ratio of 0.82 : 0.18 asteroid to comet for the Earth.

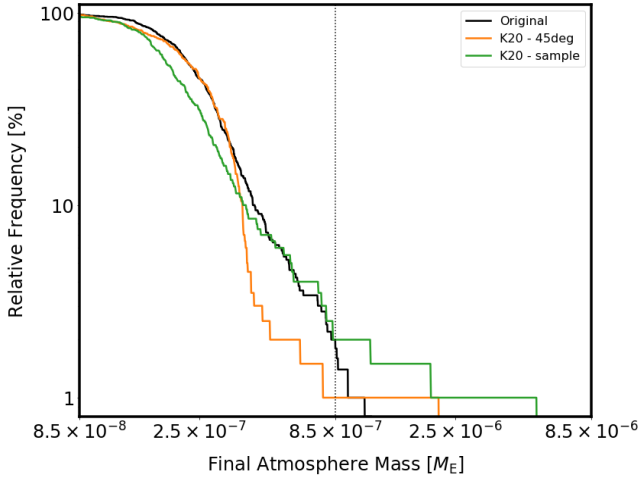

**Figure 21.** The distribution of final atmosphere masses calculated for the nominal atmosphere and impactor populations comparing the combined cratering and S15 prescription (our earlier nominal case), and the K20 prescription (both randomly sampled impact angles and an assumed impact angle of 45 deg). The S15 prescription includes the results of 500 runs, while the two K20 prescriptions include 200 runs each.

P11 find that impacts onto Earth result in the atmosphere mass remaining approximately constant through time. Considering the results of §6.1 and 6.2, we predict that asteroids should result in atmospheric growth, while comets result in loss. The assumptions we make regarding the impactor populations (discussed in §5.3) predict that there should be a marginally higher mass of impacting asteroidal material than cometary material, however the total mass of impacting material should be dominated by the left-over planetesimal population. If we assume instead that the Late Veneer was delivered by a combination of asteroids and comets only, it is possible that our mass estimates for these two populations would be more similar to those of P11 and our estimates for the total final atmosphere mass of the Earth might be in agreement. We have demonstrated that the dynamics of the impacting population can have a significant effect on the final atmosphere mass, making the difference between growth or loss, an effect that is not accounted for in P11. Furthermore, we have found that stochastic delivery of volatiles by large asteroids can cause significant variation in the final atmosphere mass, an effect that was not considered in P11, and that can result in final atmosphere masses that are an order of magnitude larger than the median result.

### 7.5.2 *de Niem et al. (2012)*

The work performed in dN12 provides an opportunity to consider how the recent advances in our understanding of the dynamical history of the Solar system affect our predictions for the evolution of Earth's atmosphere. dN12 used a similar approach to ours, stochastically sampling from distributions of impactor size and impactor velocity. However, as discussed in §6.3, they use a different prescription for the outcome of impacts, choosing a modified implementation of the model from Svetsov (2000) in comparison to our combination of Shuvalov (2009) and Schlichting et al. (2015). As we show in §7.4, the choice of impact prescription can have an effect on the atmospheric evolution predicted, however as shown in Figure 2 of dN12, the Svetsov (2000) and Shuvalov (2009) prescriptions predict broadly similar eroded atmosphere masses. The difference in predicted impactor mass accreted is also small, however as discussed in §6.3 this can cause a noticeable difference in the mass accretion predicted for large, low density impactors.

As in P11, dN12 do not consider the effect of left-over planetesimals on the atmosphere, instead taking data from simulations of an older iteration of the Nice model (Gomes et al. 2005; Morbidelli et al. 2010) then investigating a range of asteroid to comet ratios to specify the composition of the impacting population. Their distributions of velocities appear similar to our distribution for the left-over planetesimals, particularly in regards to the lack of impactors with $v_{imp} > 4v_{esc}$. This is in contrast to our predicted asteroid and comet distributions, and they furthermore do not consider the time evolution of the impact velocity distributions. The total mass that they estimate to impact the Earth ($3.3 \times 10^{-5}$ $M_\oplus$) is higher than our estimate for the total mass contained in the cometary and asteroid impactors combined ($6.1 \times 10^{-6}$ $M_\oplus$), but this neglects the significant mass contained in our population of left-over planetesimals (0.075 $M_\oplus$), which we find are the most important contribution to atmospheric evolution in the majority of our runs.

We can compare our results to those of dN12 by looking directly at the results of §6.1 and 6.2. Their assumed comet compositions lie somewhere between our "nominal" and "dry" comet populations, while their asteroids appear most similar to our "dry" asteroids. §6.2 would therefore imply that, if the impactors were entirely cometary, we might predict modest atmospheric loss while this scenario results in the largest atmosphere growth in dN12. This discrepancy is discussed in §6.3, and arises from a combination of the difference in our chosen impactor prescriptions and the on average higher velocities sampled by our distribution in comparison to that used in dN12. For the asteroids, our results predict modest atmosphere growth, although if the mass of this population is increased (as would be needed to match the total mass of impacting material used by dN12) we could conceivably recreate results similar to those of dN12. Despite these similarities, the inclusion of the population of left-over planetesimals in our work results in significantly different overall conclusions. The difference in our predictions highlights the importance of revisiting this topic in light of the advances in our understanding of the bombardment history experienced by Earth.

### 7.5.3 *Wyatt et al. (2019)*

In W19, the analytical model uses the same impact prescription as we do, albeit without the inclusion of any large impact effects. However this model assumes a simple power law size distribution, a single impactor population and a single impact velocity associated with that population, as well as neglecting any stochastic effects. The conclusions from that model applied to the Earth predict growth for impacts by asteroid like impactors, but loss for comet like impactors, in agreement with our results. This paper highlighted the sensitivity of the predictions regarding atmosphere growth and loss to the parameters of the impactor properties, in particular their velocities. Within a plausible range of impactor velocities, the W19 model can predict significant atmosphere growth for slower impact velocities and loss for faster velocities. This model cannot account for the variation in impactor velocities within a single impactor pop-ulation, which we find can be a significant source of variation in our calculated final atmosphere masses. This is illustrated by our investigation of the variation in the left-over planetesimal dynamics discussed in §6.4.1. Furthermore, we have shown that the stochastic delivery of volatiles by the largest impactors can result in significant deviation of the atmosphere mass from the "typical" evolution, which is not accounted for in the analytic model.

### 7.6 Alternative atmospheric evolutionary mechanisms

We have focused our attention in this paper on the effect of impacts on the evolution of Earth's atmosphere, neglecting a number of other effects. This is motivated by the fact that we consider only a short period of Earth's history (covering 500 Myr) during which the impact rate was high, and we might expect the effect of impacts to dominate the atmospheric evolution. However, other processes can and do influence the atmosphere; prior atmospheric evolution would have determined the atmospheric properties at the onset of the period of bombardment, while processes that occur after the end of our simulation might further alter the atmosphere, influencing how our results should be interpreted.

#### 7.6.1 Initial atmospheric conditions

As discussed in §6.6, the initial conditions for our simulations (the atmosphere that remains once the final episode of core-mantle differentiation following the Moon-forming impact has occurred) are not well constrained by observations. It has been proposed (Pepin 1991; Dauphas 2003) that the proto-Earth had a Solar-composition atmosphere that was lost through hydrodynamic escape driven by extreme-UV flux from the active young Sun, and the present day abundances and isotope ratios of the elements are a result of subsequent mantle outgassing and projectile delivery. Hydrodynamic atmosphere loss should result in fractionation of elemental isotopes, as lighter isotopes are preferentially removed with the escaping hydrogen leaving behind an atmosphere enriched in heavier isotopes relative to the mantle. At the present time, the narrative on this mostly considers the possibility that an initially large atmosphere was lost hydrodynamically and subsequently a secondary atmosphere was replenished by outgassing. This is inconclusive because while He and Ne observations agree with such a possibility (Harper & Jacobsen 1996; Ozima & Podosek 2002), Kr does not (Holland et al. 2009). Furthermore, to recreate the chondritic isotope compositions of H, C, N and Cl in the mantle (Marty 2012; Halliday 2013; Sharp & Draper 2013) requires fine tuning of the hydrodynamic loss of hydrogen which is an unlikely scenario (Schlichting & Mukhopadhyay 2018).

A further process that will have played a significant role in determining the initial conditions for our simulations are the violent large impacts predicted by dynamical simulations (Chambers & Wetherill 1998; Chambers 2001) during later stages of planet formation leading up to the Moon forming impact. These impacts are believed to result in substantial atmosphere loss (Schlichting & Mukhopadhyay 2018) and global magma oceans on Earth (Elkins-Tanton 2012), which should undergo ingassing (dissolution of volatiles from the atmosphere into the magma ocean) and outgassing (the release of volatiles from the magma ocean into the atmosphere). Outgassing of a secondary atmosphere during solidification of a magma ocean is expected to result in noble gas concentrations in the magma oceans that are fractionated according to their differing solubilities. The $^3$He/$^{22}$Ne ratios observed in plume mantle sources (that trace deep mantle with ratio $2-3$) and mid-ocean ridge basalts (that trace shallow mantle with ratio 10) are used to argue for a series of global magma oceans, since each episode can enrich the mantle by at most a factor 2 (the ratio of solubilities of He and Ne). While mantle-atmosphere exchange can explain this observation, the mantle and atmosphere ratios of Ne and Kr cannot be explained through either outgassing, or a combination of outgassing and hydrodynamic loss (Schlichting & Mukhopadhyay 2018). This leaves open the possibility that a different process has further affected the atmosphere evolution, and means we are free to consider a wide range of potential initial conditions for our simulations.

The results from §6.6 suggest that impacts are capable of removing an atmosphere with primordial composition and mass up to 10 times more massive than the present day atmosphere. Impact driven atmosphere loss would have a different isotopic signature than either hydrodynamic loss or mantle-atmosphere exchange, since it results in bulk loss of the atmosphere. Thus it might be possible to reconcile an initially high atmosphere mass with these isotopic signatures through a combination of hydrodynamic escape, ingassing, outgassing and impacts. We also note that our cratering impact prescription does not include the effect of impactor fragmentation or aerial bursts, which would increase the erosional efficiency of the smaller impactors (Shuvalov et al. 2014). This effect is predicted to be small for the conditions considered in this work, but could contribute to the removal of an initially more massive atmosphere. These effects combined could potentially effectively strip a large atmosphere, replacing it with a secondary outgassed atmosphere. More work needs to be done to understand the combined effect of these processes, and to predict what kind of secondary atmosphere would result particularly in regards to the isotopic signatures that would result from such a scenario, however we leave a detailed consideration of this as a topic for future investigation.

#### 7.6.2 Impact triggered outgassing

Our prescription for the effect of an impact on the atmosphere neglects a potentially significant effect, that of impact triggered outgassing. Our motivation for this is firstly to avoid introducing a number of unconstrained free parameters into the model and secondly because the focus of this work is the effect of impacts on the atmosphere mass directly. A sufficiently energetic impact will deliver enough energy to melt a portion of the planet's surface, from which trapped volatiles can be released into the atmosphere (through outgassing). The mass of volatiles released in this manner depends not only on the impact history, but also the volatile content of the mantle, and the properties of the planet that determine the volume of melt produced. The work of Schlichting et al. (2017) showed that impact triggered outgassing can completely negate impact driven atmosphere mass, leading to significant atmosphere growth, and so this is a mechanism worth addressing in more detail.

Using a very simplified toy model, we can estimate the mass of the Earth that would be melted as a result of impacts by the population of left-over planetesimals (we ignore the other populations since they contain negligible mass in comparison to the left-over planetesimals). An analytic expression for the volume of melt on a planet with radius $R_{\rm pl}$ and volume $V_{\rm pl}$ produced by an impactor with radius $R_{\rm imp}$ is given by Reese & Solomatov (2006),

$$V_m = \frac{1}{2} V_{\rm pl} \left(\frac{r_m}{R_{\rm pl}}\right)^3 \left(1 - \frac{3}{8}\frac{r_m}{R_{\rm pl}}\right), \qquad (28)$$

where

$$\left(\frac{r_m}{R_{\rm pl}}\right)^3 = 2\left(\frac{v_{\rm imp}}{v_{\rm imp}^{Pm}}\right)^{3/2} \left(\frac{R_{\rm imp}}{R_{\rm pl}}\right)^3. \qquad (29)$$

For dunite to melt entirely at 1 bar, the value of the critical impact velocity ($v_{\rm imp}^{Pm}$) is estimated to be 7 km s$^{-1}$ (Reese & Solomatov 2006). This gives a predicted fractional melt mass (or melting efficiency) of

$$\frac{M_m}{M_{\rm imp}} = \frac{\rho_{\rm pl}}{\rho_{\rm imp}} \left(\frac{v_{\rm imp}}{7 \text{ km s}^{-1}}\right)^{3/2}. \qquad (30)$$

For the population of left-over planetesimals (neglecting the $(1 - \frac{3}{8}\frac{r_m}{R_{\rm pl}})$ factor since this is $\approx 1$ for even the largest impactors we consider), the weighted average of the velocity distribution suggests a typical melting efficiency of $\sim 6$, allowing an estimate of the total melt mass to be calculated from the total mass of the population of left-over planetesimals. Assuming a bulk mantle volatile content in the range $0.01 - 0.15\%$ (Schlichting et al. 2017) and assuming that all the volatiles in the melt mass are outgassed, this would deliver a total mass in volatiles of $\sim 0.5 - 7 \times 10^{-5}$ M$_\oplus$ to the atmosphere. This calculated mass is insensitive to whether we use the entire population of left-over planetesimals or consider only impactors above a certain size. This is $5 - 80$ times the current atmosphere mass, and could be significantly more massive than both the predicted atmospheric erosion by left-over planetesimal impacts as well as the total mass of volatiles contained in the entire impacting population of left-over planetesimals ($2.6 \times 10^{-6}$ M$_\oplus$).

This estimated outgassed mass is an upper limit, because it assumes that each impactor melts a unique portion of the planet. It would be more reasonable to assume that impactors arriving late in the time period will remelt material that has already been melted (and outgassed the volatiles it contains) at least once. Despite this, impact triggered outgassing could overwhelm the atmospheric depletion that our results show as well as significantly alter the composition predicted for the final atmosphere. However we leave detailed inclusion of its effects to a different study since it depends on further free parameters such as the mantle volatile content.

*7.6.3 Implications for life*

Our results suggest that at the end of our simulations ($\sim 4$ Gya) the atmosphere is predominantly composed of material delivered by the population of left-over planetesimals. We have assumed that these volatiles are primarily hydrogen and carbon monoxide, with smaller contributions from carbon dioxide, nitrogen, methane, hydrogen sulfide and ammonia. A highly reduced atmosphere (methane, hydrogen and ammonia dominated) appears to be necessary for the emergence of life, however this contradicts geological evidence that the Earth's mantle has always been oxidised (carbon dioxide, water and nitrogen dominated). Zahnle et al. (2019b) propose that this disagreement could be resolved if the Late Veneer was sufficiently reducing, which is the case for dry enstatite chondrite-like impactors. However their arguments require that the Late Veneer must be delivered in a small number of massive impacts that are capable of vaporising the ocean in order to provide the high H$_2$ pressures needed to favour methane and ammonia production over carbon dioxide and nitrogen.

For our adopted size distribution the creation of a highly reducing atmosphere in their models would require either the extraction of extra reducing power from the Earth's mantle or the existence of as yet unknown catalysts. However Zahnle et al. (2019b) did not consider atmospheric erosion or volatile delivery resulting from impacts, and thus our prediction for an atmosphere dominated by volatiles delivered by the population of left-over planetesimals could potentially create a transient highly reducing atmosphere without the need for massive ocean vaporising impacts. This could potentially provide conditions conducive for the production of pre-biotic molecules without requiring impacts so violent they would wipe out any extant life.

*7.6.4 Subsequent atmospheric evolution*

The atmosphere during and immediately after the relatively intense period of bombardment we consider in our simulations is unlikely to be in thermo-chemical equilibrium, and so would be expected to continue to evolve over time following the end of our simulations. Some molecular species delivered by the impactor populations will be destroyed through photo-dissociation on various timescales, altering the chemistry of the atmospheres. Outgassing, driven by volcanism, that occurs between the end of our simulations and the present day is likely to further influence the atmosphere composition, in the manner discussed in §7.6.1. Other processes, such as the carbon-silicate cycle are known to act as feedback loops, stabilising the Earth's climate over millions of years (Walker et al. 1981). Furthermore, the emergence of life and the presence of a biosphere on Earth has significantly impacted the atmosphere, most noticeably through the Great Oxidation Event around 2.2 Gya (Lyons et al. 2014). As a consequence of all these effects, the fact that the final atmosphere compositions that we predict do not match the present day composition of the Earth's atmosphere is not necessarily a cause for concern.

## 8 CONCLUSIONS

We have presented a new numerical code to model the evolution of a terrestrial planet atmosphere due to bombardment. This code extends the analytical model for atmosphere evolution presented in Wyatt et al. (2019), accounting for the inherent stochastic nature of larger impacts and introducing an adaptive time step. The numerical code also includes a distribution of impactor velocities and multiple populations of impactors with different properties. We adopt the cratering impact prescription from Shuvalov (2009), combined with a prescription for non-local atmosphere loss caused by large impacts from Schlichting et al. (2015), but consider also an alternative prescription from Kegerreis et al. (2020). The code can successfully reproduce the atmosphere evolution for simplified impactor populations predicted by the analytical model.

To consider the evolution of Earth's atmosphere after the Moon-forming impact, we construct three populations of impactors: comets, asteroids and left-over planetesimals. The distribution of impact velocities and impact fluxes for these populations are calculated from the results of dynamical simulations of the Solar system (Nesvorný et al. 2013, 2017a; Morbidelli et al. 2018). Considering the effect on Earth's atmosphere of each of these populations individually, we find that comets in general cause the atmosphere to deplete, and this loss is greater for more massive populations, or "wetter" (more volatile rich, lower density) impactors. If the comets are assumed to be "drier" they can in some cases result in growth due to the stochastic arrival of large, slow objects. In contrast, asteroids cause atmospheric growth, and the final atmosphere mass in general increases as the assumed volatile content of the asteroids increases. Increasing the total mass of the impacting population increases the number of stochastically sampled large, slow impactors resulting in very large final atmosphere masses. Compared to the other individual populations, the stochastic effects are most obvious for the asteroid population due to the fact that for the initial atmosphere mass and impactor sizes that we assume, large (and therefore stochastic) asteroid impacts only ever lead to growth. Therefore, since large comets and left-over planetesimals deliver almost no volatiles (due to their low accretion efficiency and low volatile content respectively) only large asteroids are capable both of contributing a substantial portion of their mass to the planet, and releasing a significant fraction of that mass into the atmosphere. The left-over planetesimals always result in atmospheric erosion with the final atmosphere mass decreasing as the volatile content of the impactors is decreased, until the entire atmosphere is stripped rapidly for sufficiently "dry" impactors. Considering plausible variation in the impactor dynamics we find this can cause typical atmosphere mass loss to vary between $-72$ % and $-96$ %.

Investigating the combined effect of all three populations, our results emphasise the importance of considering stochastic events, as the relatively rare arrival of a single large impactor can have significant effects on the atmosphere mass and composition. Our results show that for identical starting conditions a wide range of outcomes is possible, with variation introduced through the uncertainty in the impactor dynamics and compositions. The results from our nominal case show modest atmospheric loss, with a median final atmosphere mass of $0.24 \times 10^{-6}$ $M_\oplus$. The sampling of a large, slow asteroid can result in significantly higher final atmosphere masses than the median value. We find that the final atmosphere is in general dominated by material delivered by the population of left-over planetesimals, with smaller primary components and even smaller contributions from asteroids and comets. The cometary fraction we find, typically $< 1$ %, is consistent with observational constraints.

The initial mass and composition assumed for the Earth's atmosphere appears to make relatively little difference to the final outcome, however we can rule out an initial atmosphere mass significantly greater than the present mass as the material delivered in the Late Veneer is not capable of sufficient erosion. The exception to this would be if the initially massive atmosphere was primordial, as atmospheres with lower $\mu_0$ appear easier to remove.

Further work needs to be done to better understand the isotopic signatures of the different atmosphere evolution histories we predict, and to understand how these relate to observational constraints on the Earth. In addition, predicting the evolution of atmospheres on Venus and Mars, using appropriate dynamics for the impactor populations is also worth exploring.

**ACKNOWLEDGEMENTS**

C.A.S. acknowledges the support of STFC via the award of a DTP Ph.D. studentship.

**DATA AVAILABILITY**

Data available on request.

This paper has been typeset from a T<sub>E</sub>X/L<sup>A</sup>T<sub>E</sub>X file prepared by the author.